\documentclass[aip,graphicx]{revtex4-1}
\usepackage{graphicx}
\usepackage{gensymb}
\usepackage{epsfig}
\usepackage{color}
\usepackage{epstopdf}
\usepackage{amsmath}
\usepackage{caption}
\usepackage{subcaption}
\usepackage{url}

\newcommand{\grl}{    {Geophys. Res. Lett.}}
\newcommand{\jgr}{    {J. Geophys. Res.}}
\newcommand{\ssr}{    {Space Sci. Rev.}}

\newcommand{\aap}{    { Astronomy and Astrophysics}}

\newcommand{\apj}{ {Astrophys. J. }}

\newcommand{\nat}{ {Nature }}

\newcommand{\prl}{ {Phys. Rev. Lett.}}
\newcommand{\pre}{ {Phys. Rev. E}}
\newcommand{\mnras}{ {Monthly Notices of the Royal Astronomical Society}}
\newcommand{\apjl}{ {Astrophys. J. Lett. }}

\begin{document}


\title{On application of stochastic differential equations for simulation of nonlinear wave-particle resonant interactions} 



\author{A. S. Lukin}
\affiliation{
Space Research Institute of the Russian Academy of Sciences (IKI), 84/32 Profsoyuznaya Str, Moscow, Russia, 117997
}%
\email{as.lukin.phys@gmail.com}
\affiliation{Faculty of Physics, National Research University Higher School of Economics, 21/4 Staraya Basmannaya Ulitsa, Moscow, Russia, 105066}

\author{A. V. Artemyev}
\affiliation{
Department of Earth, Planetary, and Space Sciences, University of California, 595 Charles E Young Dr E, Los Angeles, CA, California, USA, 90095
}%
 \email{aartemyev@igpp.ucla.edu}
\affiliation{
Space Research Institute of the Russian Academy of Sciences (IKI), 84/32 Profsoyuznaya Str, Moscow, Russia, 117997
}%

\author{A. A. Petrukovich}
\affiliation{
Space Research Institute of the Russian Academy of Sciences (IKI), 84/32 Profsoyuznaya Str, Moscow, Russia, 117997
}%
 \email{a.petrukovich@cosmos.ru}


\date{\today}

\begin{abstract}
Long-term simulations of energetic electron fluxes in many space plasma systems require accounting for two groups of processes with well separated time-scales: microphysics of electron resonant scattering by electromagnetic waves and electron adiabatic heating/transport by mesoscale plasma flows. Examples of such systems are Earth’s radiation belts and Earth’s bow shock, where ion-scale plasma injections and cross-shock electric fields determine the general electron energization, whereas electron scattering by waves relax anisotropy of electron distributions and produces small populations of high-energy electrons. The applicability of stochastic differential equations is a promising approach for including effects of resonant wave-particle interaction into codes of electron tracing in global models. This study is devoted to test of such equations for systems with nondiffusive wave-particle interactions, i.e. systems with nonlinear effects of phase trapping and bunching. We consider electron resonances with intense electrostatic whistler-mode waves often observed in the Earth’s radiation belts. We demonstrate that nonlinear resonant effects can be described by stochastic differential equations with the non-Gaussian probability distribution of random variations of electron energies.
\end{abstract}


\maketitle

\section{Introduction}
The theoretical description of wave-particle interactions in space collisionless plasma is traditionally based on the quasi-linear approach proposed for a broadband low amplitude wave ensemble in Refs. \onlinecite{Drummond&Pines62, Vedenov62} and further developed in \onlinecite{Kennel&Engelmann66,Lerche68,Andronov&Trakhtengerts64}. This approach provides quite powerful tools to model evolution of wave spectra and resonant particle distributions for various space plasma systems (see, e.g., reviews in Refs. \onlinecite{Bykov&Toptygin93,Thorne10:GRL, bookSchulz&anzerotti74, bookLyons&Williams, Shprits08:JASTP_local}). Quasi-linear models describe well relativistic electron acceleration in the Earth’s radiation belts\cite{Horne05Nature,Thorne13:nature,Allison&Shprits20}, relaxation of anisotropy-driven instabilities in the solar wind \cite{Shaaban19}, and energetic particle generation at astrophysical shock waves \cite{Schure12:diffusive_shock_acceleration,Bykov17}. Being developed for broadband ensemble of plane waves, this approach has been successfully generalized to systems with narrow band wave spectrum (when the background magnetic field gradients may destroy wave phase correlation, see Refs. \onlinecite{Solovev&Shkliar86,Albert10}) and systems with particle scattering by nonlinear (but low amplitude) solitary waves \cite{Vasko17:diffusion, Vasko18:pop, Shen20:tds}. However, more precise wave measurements in the new spacecraft missions posses new requirements to wave-particle interaction models (see discussion in Refs. \onlinecite{Albert13:AGU,Mourenas18:jgr, Zhang18:jgr:intensewaves,Zhang20:grl:phase, Li&Hudson19}), and such requirements cannot be addressed within the quasi-linear models that all are based on Fokker-Plank diffusion equation \cite{bookTrakhtengerts&Rycroft08,bookSchulz&anzerotti74}. A good example is the set of quasi-linear models describing electron resonant interaction with the whistler-mode waves, that are widely observed on shock waves \cite{Zhang99:whistlersBS}, solar wind\cite{Tong19:ApJ}, and planetary magnetospheres \cite{Agapitov13:jgr, Li12, Menietti12}. Very high amplitudes of these waves \cite{Cattell08, Cattell11:Wilson,Oka17, Hull12,Zhang19:grl:whistlers} and their association with large-scale mesoscale dynamics (e.g., transient plasma flows \cite{LeContel09, Breuillard16,Zhang18:whistlers&injections} and low-frequency large-amplitude compression perturbations \cite{Oka19, Zhang19:jgr:modulation}) require development of new approaches that would significantly extend the quasi-linear one.

There are two groups of models that pretend to resolve such issues of description electron resonances with whistler-mode waves. Models of the first group are focused on inclusion of high wave intensity that drives the nonlinear wave-particle interactions, e.g. phase trapping and phase bunching (see reviews in Refs. \onlinecite{Karpman74:ssr, Shklyar09:review,Albert13:AGU,Artemyev18:cnsns}). These models are based on a comprehensive analysis of individual resonant orbits \cite{Shklyar81,Shklyar&Zimbardo14,Albert93,Albert01,Demekhov06,Demekhov09,Artemyev12:pop:nondiffusion,Artemyev14:pop} and following generalization of this analysis for construction of integral operators supplementing the Fokker-Plank equation \cite{Furuya08,Artemyev14:grl:fast_transport,Omura15}. Such operators can be constructed numerically (see examples in Refs. \onlinecite{Hsieh&Omura17:radio_science, Hsieh&Omura17, Zheng19:emic,Artemyev19:cnsns}) or analytically (see examples in Refs. \onlinecite{Artemyev16:pop:letter, Artemyev18:jpp,Vainchtein18:jgr}). This approach competes with the massive test particle simulations \cite{Agapitov15:grl:acceleration,Camporeale&Zimbardo15,Gan20:grl,Gan20:grl:II, Allanson19, Allanson20} and allows including nonlinear resonant effects into long-term simulations of electron distribution functions. However, such generalized Fokker-Plank equations do not include any large-scale and mesoscale transient variations of electron distributions that are extremely important for radiation belts (e.g., electron injections and adiabatic acceleration \cite{Gabrielse14, Gabrielse19, Birn14, Turner16}, non-diffusive scattering on magnetic field gradients \cite{Eshetu18}, and losses through the Earth magnetosphere boundary –- magnetopause \cite{Sorathia18}) and shock waves (e.g., adiabatic heating \cite{Schwartz11:VladimirShock} and scattering on electric field gradients \cite{Balikhin93, Gedalin20}).

Models of the second group describe electron dynamics in large-scale electromagnetic fields derived from global MHD simulations (mostly considered for radiation belts\cite{Elkington04, Hudson12:simulation,Hudson15, Maha13:lsk}) and global hybrid simulations (mostly used for Earth's bow shock \cite{Liu19:foreshock}). The basic realization of this approach allows to include radial transport (as MHD simulations resolve well ultra-low-frequency waves) and nonadiabatic losses through the magnetopause \cite{Sorathia17, Sorathia18}. Moreover, the most modern adaptation of such models mimic effects of quasi-linear electron scattering \cite{Elkington18:agu,Elkington19:agu}. Such modification has been done via substituting the integration of electron trajectories by solving stochastic differential equations \cite{Tao08:stochastic,Zheng14:stochastic}. Although, this merging of electron transport in realistic magnetic fields (given by MHD simulations) and scattering by realistic waves (given by statistical spacecraft observations) is very promising, it still ignores all nonlinear effects of wave-particle resonant interaction. Moreover, simulations of electron pitch-angle and energy diffusion within such an approach do not account for magnetic field evolution and based on dipole field model, i.e. effect of electron movement into/out from resonances by adiabatic energy change is omitted.

Inclusion of resonant scattering as a perturbation of electron orbits tracing in the global MHD/hybrid models seems to be the most perspective direction in simulations of electron fluxes in radiation belts and at the bow shock. However, the standard approach of stochastic differential equations \cite{Tao08:stochastic} has been developed to describe diffusive-like scattering, and a further modification of this approach is needed to incorporate nonlinear effects. This study aims to consider and test one of the possible solutions of such incorporation. We focus on a particular system with electron resonant interaction with electrostatic mode of very oblique whistler waves \cite{Agapitov15:grl:acceleration, Artemyev16:ssr}. For this system we demonstrate that the classical stochastic differential equations with the Wiener probability distribution can be generalized to include probabilities of phase trapping and phase bunching. Set of verifications with test particle simulations show that such generalization works well.

\section{Wave field model}
We investigate electron resonant interaction with quasi-electrostatic whistler-mode waves widely observed in the Earth's radiation belts \cite{Agapitov13:jgr, Cattell15} and contributing significantly to electron acceleration \cite{Agapitov14:jgr:acceleration,Agapitov15:grl:acceleration}. The Landau resonance with such waves resembles electron resonance with electrostatic Langmuir waves \cite{Karpman75, Shklyar81, Artemyev17:pre, Tobita&Omura18}: this resonance does not change magnetic moment $\mu$, and thus wave-particle interaction can be described by $
1{\textstyle{1 \over 2}}$ degrees of freedom Hamiltonian \cite{Artemyev12:pop:nondiffusion}:
\begin{equation}\label{H_dim}
    H=\frac{1}{2m}p_{\parallel}^2+\mu B(s) - e\Phi(s) \sin{\varphi}
\end{equation}
where $m$, $-e$ are electron mass and change, $(s,p_\parallel)$ are field-aligned coordinate and momentum (the pair of conjugated variable), $\mu=H\sin^2\alpha/B(s)$ is the magnetic moment of particles with $\alpha$ pitch-angle (through the paper we use the equatorial pitch-angle $\alpha_{eq}$ defined at $B(0)$), $\Phi$ is the wave scalar potential, and $\varphi$ is the wave phase defined as $\dot\varphi=k_\parallel\dot{s}-\omega$. The parallel wave number $k=k_\parallel(s,\omega)$ is determined from the cold plasma dispersion relation for constant wave frequency $\omega$:
\begin{equation}
k_\parallel  = \frac{{\omega _{pe} }}{c}\left( {\frac{{\Omega _{ce}(s) \cos \theta }}{\omega } - 1} \right)^{ - 1/2}
\end{equation}
where $\Omega_{ce}=eB(s)/mc$. We use constant plasma frequency $\omega_{pe}$ determined from the empirical density model\cite{Sheeley01} as a function of the radial distance from the Earth ($R=R_EL$ and $R_E$ is the Earth radius). For wave normal angle $\theta$ we use the model proposed in Ref. \onlinecite{Agapitov14:jgr:acceleration}: $\cos\theta=\cos\left(\theta_r-\Delta\theta\right)$ with $\cos\theta_r=\omega/\Omega_{ce}(s)$ (resonant cone angle\cite{bookStix62}) and $\Delta\theta=const$. This model is based on spacecraft observations of quasi-electrostatic whistler-mode waves propagating with $\theta$ of few degrees away from $\theta_r$ (see Refs. \onlinecite{Agapitov14:jgr:acceleration, Agapitov15:grl:acceleration, Li16:statistics, Li16}).

For background magnetic field models we use $B(s)=B_{eq}+(s/R)^2+a\cdot(s/R)^4$ with the spatial scale $R=R_EL$ of the Earth's dipole magnetic (i.e. the field curvature radius is $R/3$). Term $\sim s^4$ does change significantly the resonant electron interaction with whistler waves, but this term makes electron bounce period a function of electron energy, i.e. this term makes electron bounce oscillations nonlinear and this is crucially important for destruction of the phase coherence in realistic plasma systems (see, e.g., discussion in Refs. \onlinecite{Karpman74:ssr, Albert10,Artemyev17:pre}).

We introduce dimensionless variables $s\to s/R$, $p_\parallel\to p_\parallel/\sqrt{h_0m}$, $t\to t\sqrt{h_0/m}/R$, $H\to H/h_0$, $\mu\to\mu/h_0B(0)$, and parameters $\omega_m=\omega/\Omega_{ce}(0)$, $\varepsilon u(s)=e\Phi(s)/h_0$, $k=k_\parallel/k_0$, where $h_0=10$ keV, $k_0=(\omega_{pe}/\Omega_{ce}(0))\eta/\sqrt{\omega_{m}^{-1}-1}$, $\eta=R\Omega_{ce}(0)/c$, $\varepsilon = e\max\Phi /h_0=eE_0/k_0h_0$, and $E_0$ is the wave parallel eclectic field amplitude. Through the paper we mainly use $L=6$ (the outer radiation belt) and $\omega_m=0.35$ (typical whistler-mode frequency\cite{Meredith12}).

The dimensionless function $u(s)$ describes the wave-field distribution along magnetic field-lines. For $u(s)$ we consider the model based on spacecarft observations of whistler-mode waves generated around the equatorial plane $s=0$, amplified during the propagation, and finally damped at high latitudes\cite{Agapitov13:jgr,Agapitov18:jgr}:
\begin{equation}
    u(s)=0.25\left(\tanh\left(10\cdot\left(s-0.27\right)\right)+1\right)\left(\tanh\left(3-s\right)+1\right)
\end{equation}
This $u(s)$ describes a nonzero wave intensity only within a limited $s$ range. Waves are generated and propagate equally in both directions away form $s=0$, and thus $s>0$ is sufficiently to be considered for resonant interactions, because resonances at $s<0$ have the same properties (i.e., we increase time between two resonant interactions from the half of the bounce period to one bounce period).

The dimensionless Hamiltonian can be written as
\begin{equation}\label{H_norm}
    H=\frac{1}{2}p^2+\mu b(s)+\varepsilon u(s)\sin(\varphi)
\end{equation}


The full set of equations required for particle trajectory calculation is
\begin{eqnarray}\label{H_eq_norm}
    \dot{p_{\parallel}}&=&-\mu\frac{db}{ds}-\varepsilon \left(\frac{du}{ds} \sin{\varphi}+k_0 K(s)u(s)\cos(\varphi)\right) \nonumber\\
    \dot{s}&=&p_{\parallel}, \;\;\;
    \dot{\varphi}=k_0K(s)p_{\parallel}-\eta \omega_m
\end{eqnarray}

\begin{figure}
\centering
\includegraphics[width=0.75\textwidth]{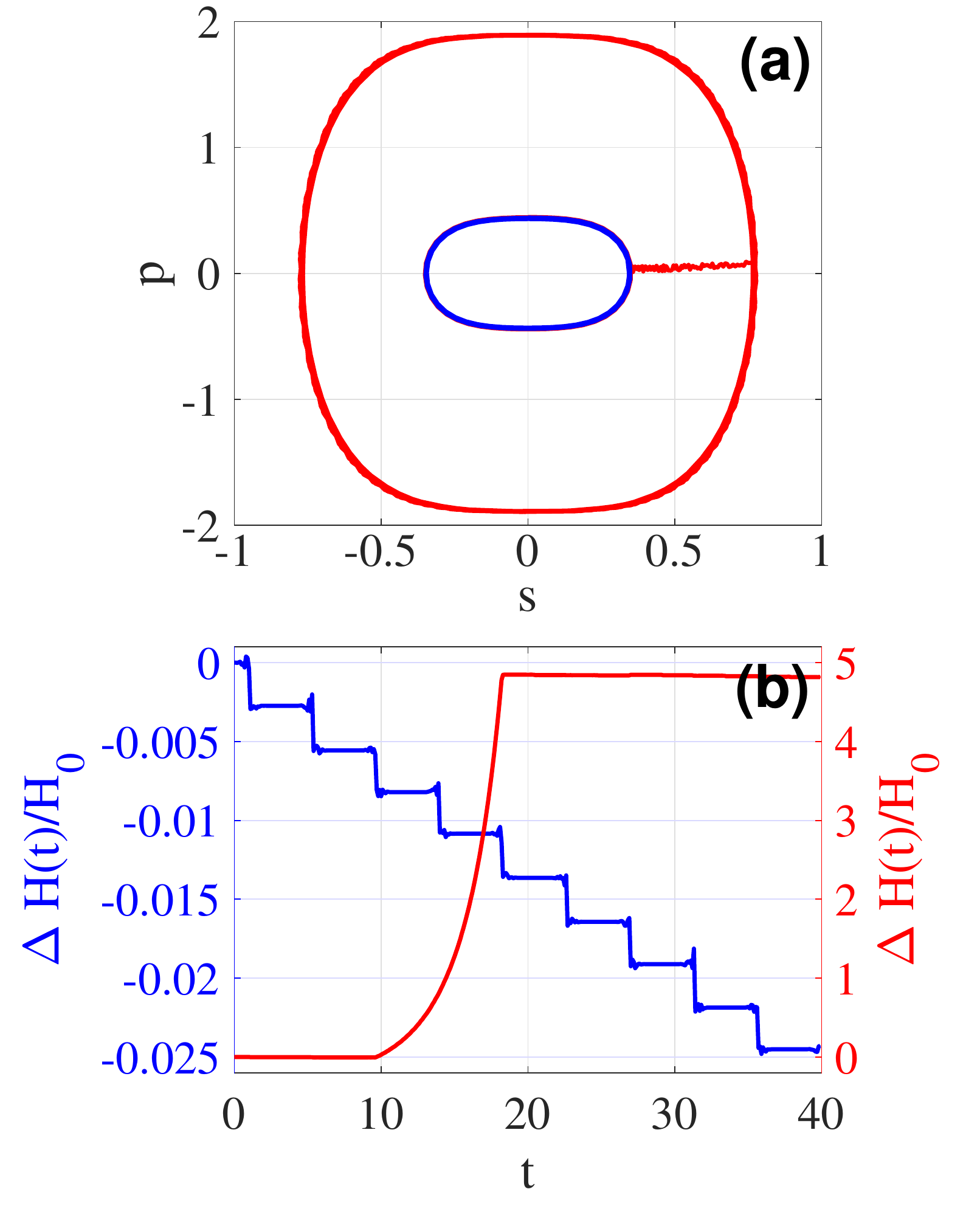}
\caption{Two examples of electron trajectories and energy jumps for the initial energy $H_0=0.35$ and two different initial phases. System parameters are: $L=6$, $\mu=0.25$, $\omega_m=0.35$, and $E_0=5$ mV/m. (a) trajectories in $(s,p_\parallel)$ plane, red and blue curves show trajectories of trapped and scattered particles; (b) red line shows energy profile for the trapped electron, while blue one shows energy profile for electrons experiencing many scatterings.}
\label{fig1}
\end{figure}

Figure \ref{fig1} shows typical particle trajectories  and energy changes  obtained from numerical integration of Eqs. (\ref{H_eq_norm}) for initial energy $H_0=0.35$ and two initial phases. There are two main resonant phenomena: phase trapping and phase bunching (scattering). Far from the resonance $\dot\varphi=0$ electrons are bouncing in $B(s)$ field (see closed trajectories in the panel (a)). During such periodical motion electrons cross the resonant $s_R$ (determined from $k_0K(s)p_\parallel-\eta \omega_m=0$ with $p_\parallel=\sqrt{2H-2\mu b(s)}$), and for $s_R>0$ (where wave field is not equal to zero) the wave can trap or scatter electrons. The scattering results in small energy change, and scatted electrons stay almost on the same orbit in $(s,p_\parallel)$ plane (see blue curves in the panels (a) and (b)). If the resonant wave-particle  interaction is nonlinear, most of the scattered particles should lose the energy (for Landau resonance), but there is an spread around such mean energy change due to a finite resonance width\cite{Karimabadi90:waves, Vainchtein17}. The phase trapping results in significant energy change (acceleration for the Landau resonance\cite{Artemyev12:pop:nondiffusion}; see red curve in panel (b)): electron is transported by the wave from one orbit in the $(s,p_\parallel)$ plane to  another orbit (see red curve in the panel (a)). For fixed wave characteristics, the total energy gain of trapped electrons depends on the initial energy and $\mu$ (those determine the resonant $s_R$). However, the finite resonance width and peculiarities of the trapped electron dynamics result in the spread of the gained energy \cite{Vainchtein17}.


\section{Statistics of energy jumps}
In a small wave amplitude limit (i.e., in absence of nonlinear effects of wave-particle interaction) and for sufficiently strong phase stochastization (i.e., when wideness of wave ensemble\cite{Shapiro&Sagdeev97} or background magnetic field inhomogeneity\cite{Albert10} destroy the correlation between resonances), the Fokker-Plank equation would describe evolution of the particle distribution. Due to conservation of magnetic moment in Landau resonance, such equation can be written for a 1D $f(H)$ distribution of electrons having the same magnetic moment:
 \begin{equation}
    \frac{\partial f}{\partial t} = af-\frac{\partial}{\partial H} \left(\nu f\right) + \frac{\partial^2}{\partial H^2} \left(Df\right)\label{eq:diff}
\end{equation}
where $a(H)$, $\nu(H)$, and $D(H)$ are source, convection and diffusion coefficients that satisfy relation $\nu=\partial D/\partial H$ and $a=\partial\nu/\partial H$ for the classical quasi-linear diffusion equation\cite{bookSchulz&anzerotti74}:
\begin{equation}
    \frac{\partial f}{\partial t} = \frac{\partial}{\partial H} \left(D\frac{\partial f}{\partial H} \right)
\end{equation}
Characteristics of Eq. (\ref{eq:diff}) are Ito equations (stochastic differential equations, SDE)
\begin{equation}
    H(t+dt)=H(t)+\nu\left(H(t),t\right)dt+\sigma\left(H(t),t\right)dW_t
\end{equation}
where $\sigma\left(H,t\right)=\sqrt{2D\left(H,t\right)}$ $dW_t=\beta \sqrt{dt}$, and $\beta \sim N(0,1)$ is a random value with the Gaussian distribution having zero mean and unity dispersion.

The basic requirements of Fokker-Planck equation (or corresponding Ito equations)  applicability is the smallness of $\Delta H$ jumps. This requirement does not work for the nonlinear wave-particle interaction associated with large $\Delta H$ change even for a single resonant interaction (see Fig. \ref{fig1}). However, such nonlinear resonant interaction is also probabilistic process \cite{Artemyev17:pre, Artemyev20:pop}, and thus we can generalize Ito equation to include such large $\Delta H$.

\begin{figure}
\centering
\includegraphics[width=0.8\textwidth]{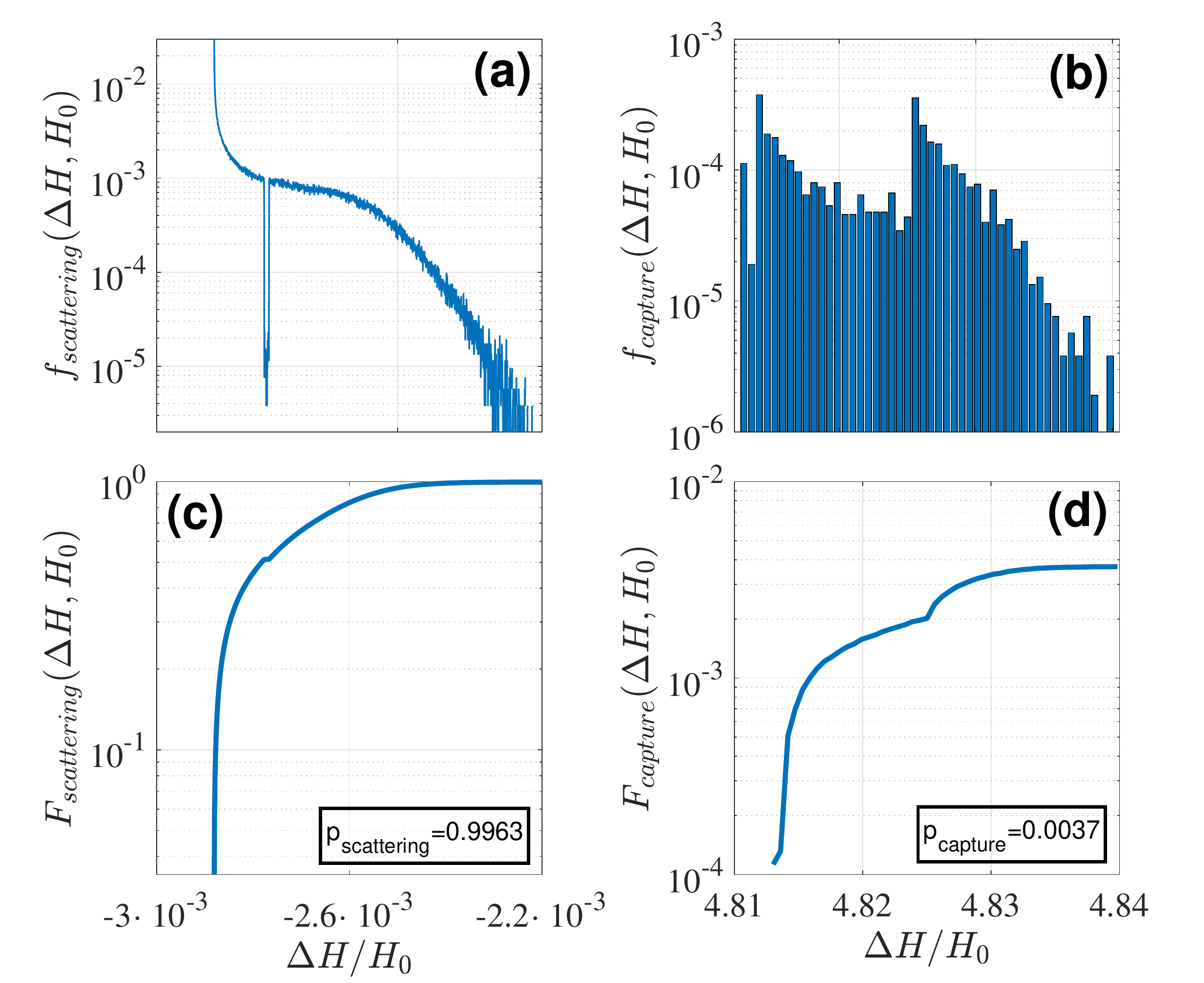}
\caption{Typical examples of $\Delta H$ distributions for electrons experienced phase bunching (a) and  phase trapping (b). Panels (c) and (d) show corresponding cumulative functions. System parameters are the same as in Fig. \ref{fig1}}
\label{fig2}
\end{figure}

Figure \ref{fig2} shows typical example of $\Delta H$ distribution after a single resonant interaction (i.e. one bounce period) for the same system parameters as in Fig. \ref{fig1}. We distinguish two processes: the scattering and the phase trapping. Panel (a) shows $\Delta H$ distribution for particles experienced scattering: energy changes of such particles are sufficiently small $\Delta H/H_0 \ll 1$, i.e. there are particles gained and lost energies due to scattering. For nonlinear regime of wave-particle interaction this scattering is dominated by the phase bunching, when most of electrons lose their energy (see the strong peak at $\Delta H/H_0<0$ in panel (a)). Panel (b) shows $\Delta H$ distribution for phase trapped particles: all particles gain a significant energy with $\Delta H/H_0 \sim 4.8$. The spread of $\Delta H$ around  $\Delta H/H_0 \sim 4.8$ is due to effect of a finite resonance width. Our basic idea is to evaluate such $\Delta H$ distributions for a wide range of initial energies and then use them instead of the $N(0,1)$ normal distribution in the Ito equations.

Using distributions from Fig. \ref{fig2}, we consider wave-particle interaction as a probabilistic process and then introduce probability for both phase bunching and phase trapping as:
\begin{equation}
    p_{i}=\lim_{N\to \infty} \frac{N_{i}}{N}
\end{equation}
where $i=(scattering, \; capture)$ and $N$ is a total number of particles. The main idea behind this approach is that the energy change (i.e., due to trapping or scattering) is determined by the initial phase value far from the resonance. However, phase is fast oscillating variable and even small change $\sim O(1/\eta)$ of it's initial value would result in substantially different phase values at the resonance. Therefore, one can consider the distribution of initial phases as random and apply approaches of the probability theory to describe resonant interaction of the large particle ensemble (see detains in Refs. \onlinecite{Itin00, Artemyev20:pop}).

For fixed wave characteristics, distributions of $\Delta H$ (as shown in Fig. \ref{fig2}) can be evaluated for the given grid of initial energies $H_0$. Number of test particles required to reproduce all important details of such distributions depends on the system (wave and background field characteristics) characteristics. For example, in the system under consideration the trapping probability for sufficiently wide $H_0$ range is above $10^{-4}$, and to describe so small probability we would need at least $10^4-10^5$ test trajectories. Through the paper all $\Delta H$ distributions (for given $H_0$) are constructed from numerical integration of $6.5\cdot10^4$ trajectories, and we have examined that the increase of this number does not change results. Although the total number of test trajectories (for all $H_0$) well exceeds $10^6$, their integration is required only for an interval less than half of the bounce period (interval of the resonant interaction), whereas all further simulations are based on the constructed $\Delta H$ distributions.


We introduce the two-step approach for simulation of the electron distribution evolution based on the stochastic differential equations. The first step consists in evaluation of $\Delta H$ distributions. For fixed set of system parameters we define a range of energies and introduce the grid of initial energies $H_0$. Then we numerically integrate necessary number of test trajectories (on a half of bounce period intervals) for each $H_0$ value from the determined energy grid. For each $H_0$ we construct two probability densities $f_{scattering}$ and $f_{capture}$  for phase bunching and phase trapping, and recalculate them to  cumulative distribution functions $F_{scattering}(\Delta H)$ and $F_{capture}(\Delta H)$ (see in panels (c) and (d) in figure \ref{fig2}). Then we use the interpolated inverse functions $\Delta H = F^{-1}(F,H_0)$ where $F \in [0,1]$ is uniformly distributed random variable.

The second step consists in application of $\Delta H(F, H_0)$ functions for numerical integration of a large number of stochastic differential equations. We introduce an array of initial energies $H_0$ (stochastic trajectories) satisfying the initial energy distribution. Then we define the number of bounce periods $N_{bp}$ required to trace electron dynamics for time interval of interest. For each step $i \in [1,N_{bp}]$ we generate two uniform random numbers $U \in [0,1]$ and $F \in [0,1]$ and recalculate the energy for each stochastic trajectory as $H_{i}=H_{i-1}+\Delta H(F,H_{i-1})$, where $\Delta H(F,H_{i-1})$ is given by equation
\begin{eqnarray*}
{\rm if}\;\;\; U\le p_{scattering}(H_{i-1}) \;\;\;{\rm then}\;\;\; \Delta H=F^{-1}_{scattering}\\
{\rm if}\;\;\;  U>p_{scattering}(H_{i-1}) \;\;\;{\rm then}\;\;\; \Delta H=F^{-1}_{capture}
\end{eqnarray*}
The time along trajectory is traced as $t_i=t_{i-1}+T(H_{i-1})$ where $T(H_{i-1})$ is a bounce period of particles with energy $H_{i-1}$. Results of tracing of large trajectory ensemble are used to construct the energy distribution of electrons at any moment of time. Although we consider only one Landau resonance per bounce period, the same scheme can be generalized on arbitrary number of both Landau and cyclotron resonances, because all these resonances will change electron energy/pitch-angle in succession.



\begin{figure*}
\centering
\includegraphics[width=0.95\textwidth]{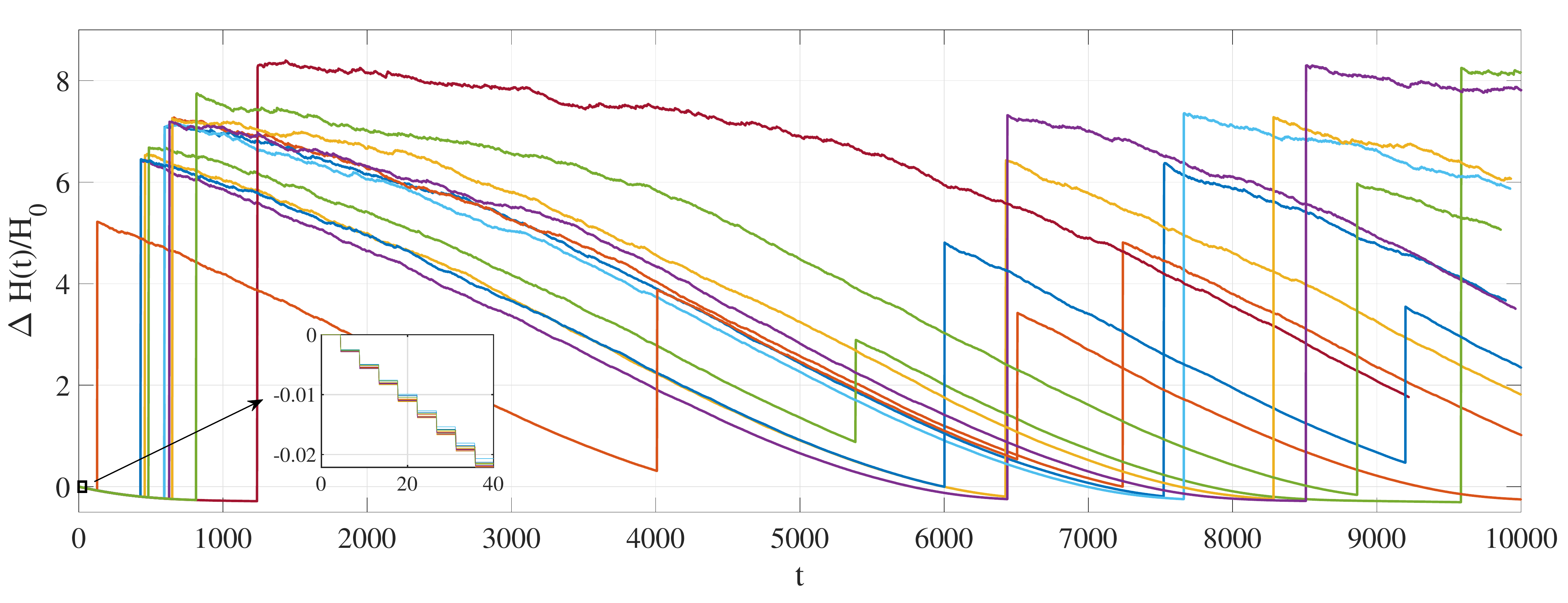}
\caption{Twelve examples of $\Delta H(t)/H_0$ obtained from the SDE approach for $4\cdot10^3$ resonant interactions for the same system parameters as in Fig. \ref{fig1}. Small panel zooms the same time interval as one shown in Fig. \ref{fig1}.}
\label{fig3}
\end{figure*}

Figure \ref{fig3} shows several examples of applications of the SDE approach for energy change in long-term simulations of electron trajectories. Both effects of phase trapping and scattering are well seen: electrons are generally scattered with the energy lost (phase bunching) and rarely trapped with the energy gain. Such trapping is possible only for low energies, and after acceleration electrons slowly drift to lower energies due to the scattering. Then the trapping repeats and electrons gain energy. This trapping-loss cycle (with some variations of energies where trapping occurs due to the probabilistic nature of trapping) is the typical for long-term electron dynamics in systems with the nonlinear resonant interaction \cite{Artemyev18:jpp}.

\section{Evolution of particle distribution dynamics}
The main advantage of using SDE equations instead of full integration of particle trajectories is the computation efficiency of this approach. SDE do not trace the longest adiabatic particle motion and include only resonant effects. Thus, the main computational resources are used for accurate evaluation of $\Delta H$ distributions (the first step of this approach). Although we evaluate these distribution numerically (this is the widespread method for different schemes of tracing distributions of electrons nonlinearly interacting with waves, see Refs. \onlinecite{Omura15,Hsieh&Omura17:radio_science,Artemyev19:cnsns}), the simplified version (without the effect of a finite resonance width) of such distributions can be evaluated analytically\cite{Artemyev20:jpp:arxiv}.  

To verify the proposed SDE approach of evaluation of the electron distribution function dynamics, we compare results obtained with SDE and with the full integration of an ensemble of test particle trajectories. Number of trajectories used to describe the electron distribution evolution are $1.6\times10^4$ and $6.5\times10^4$ for test particles and SDE method, respectively. The energy grid for calculation of the initial $\Delta H$ distributions for the SDE approach consists of $200$ logarithmically spaced bins in $H\in[H_m,20]$. We use the initial energy distribution given by the power law function $ f(H)\sim (H-H_m)^2/H^5$ with $H_m=0.25$.

\begin{figure}
\centering
\includegraphics[width=0.65\textwidth]{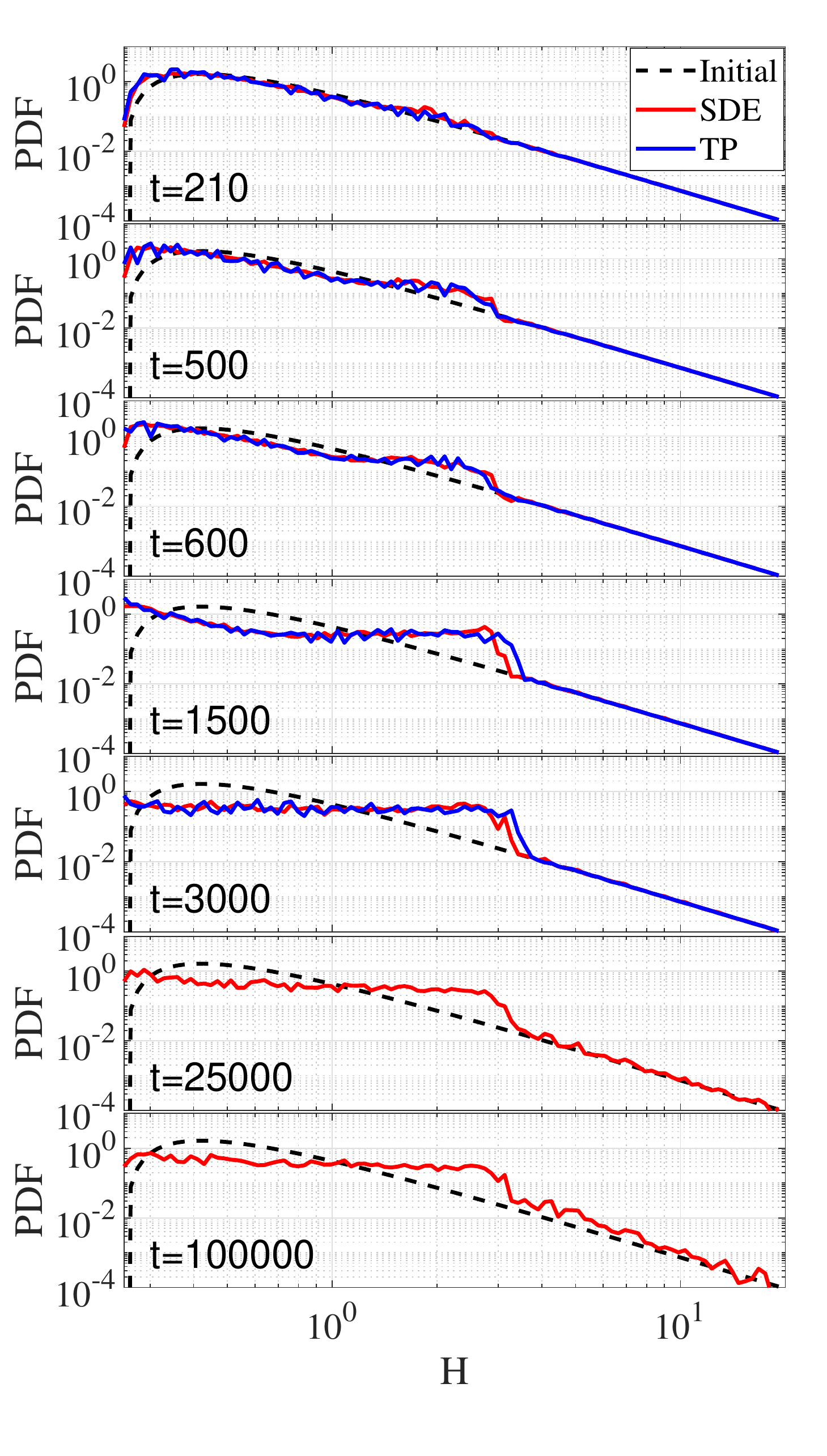}
\caption{Electron energy distribution for seven moments of time; dashed lines show the initial distribution, blue and red solid lines show test particle and SDE results, respectively. Wave amplitude is $E_w=5$}
\label{fig4}
\end{figure}

\begin{figure}
\centering
\includegraphics[width=0.65\textwidth]{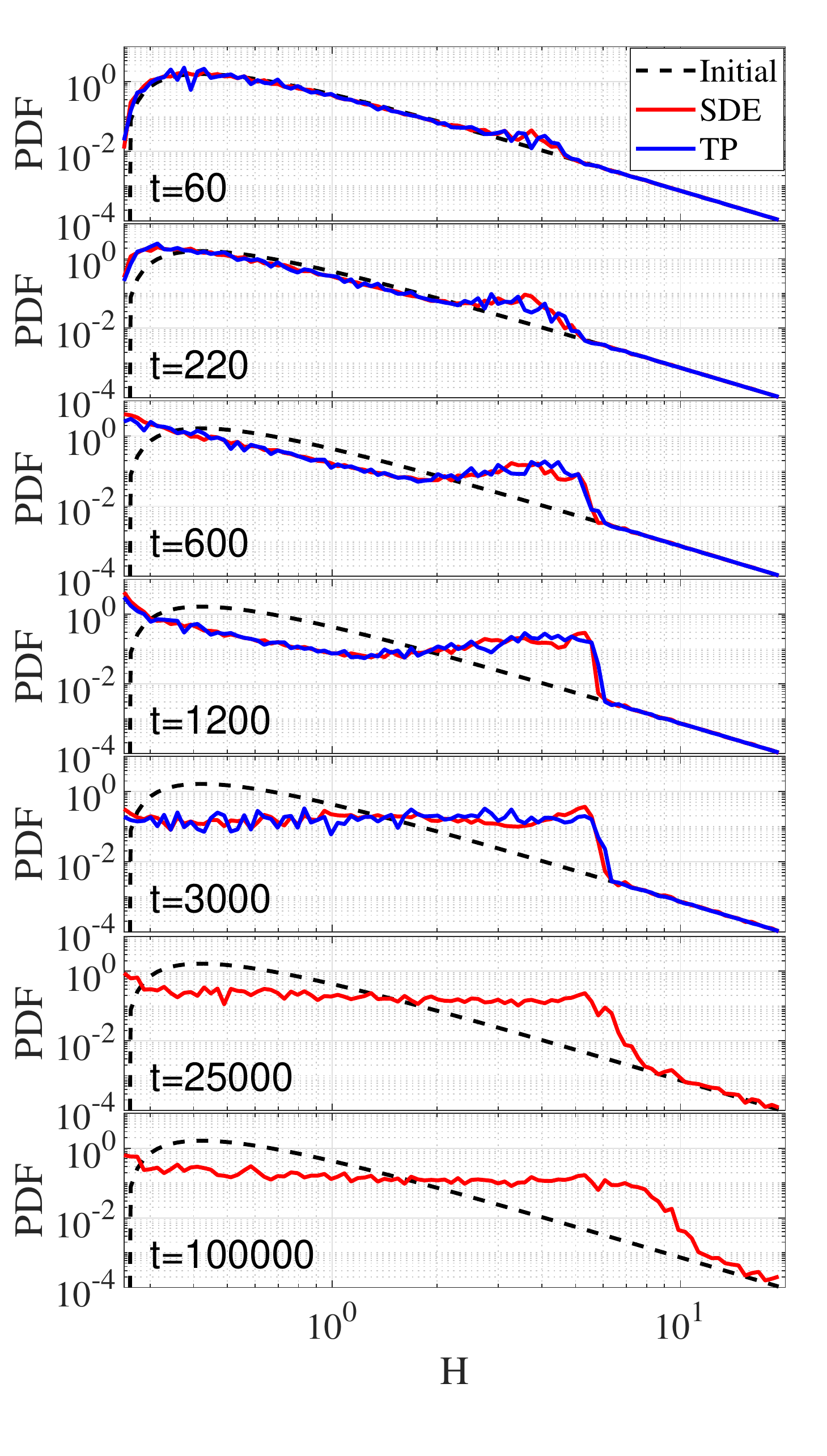}
\caption{The same as in Fig. \ref{fig4}, but for wave amplitude $E_w=10$}
\label{fig5}
\end{figure}

Figures \ref{fig4}-\ref{fig5} compares results of test particle simulations and the SDE approach application for two different wave electric field magnitudes and five time moments. There is quite good agreement between two methods that differ only at the beginning (while the fully random wave-particle interaction has not been established) and for the largest energies (where particle statistics is not sufficiently good). For $t=500$ there is the accelerated population at $H\sim 8$, and this population is generated by the phase trapping mechanism. After formation, this population starts moving to smaller energies due to the phase bunching, whereas newly trapped particles arrive to $H\sim 8$. This is the typical evolution of electron distribution driven by the nonlinear wave-particle interaction \cite{Artemyev19:pd}. 

At large energies the SDE approach shows small amplitude oscillations. For this energy range wave-particle interactions do not result in any significant energy change, and thus numerical errors of $\Delta H$ determination dominate the distribution function dynamics. There $\Delta H$ distributions are almost symmetrical relative to $\Delta H=0$ with the mean values $\langle\Delta H\rangle \sim 0$, but  $\langle\Delta H\rangle(H_0)$ may fluctuate around zero for different $H_0$. Such fluctuations lead to grouping of particles around energies with $\langle\Delta H\rangle \sim 0$. This technical issue can be resolved either by reduction of the time step $dt$ for $\Delta H$ distribution evaluation, or by increasing the detailing of the energy binning.

\begin{figure}
\centering
\includegraphics[width=0.65\textwidth]{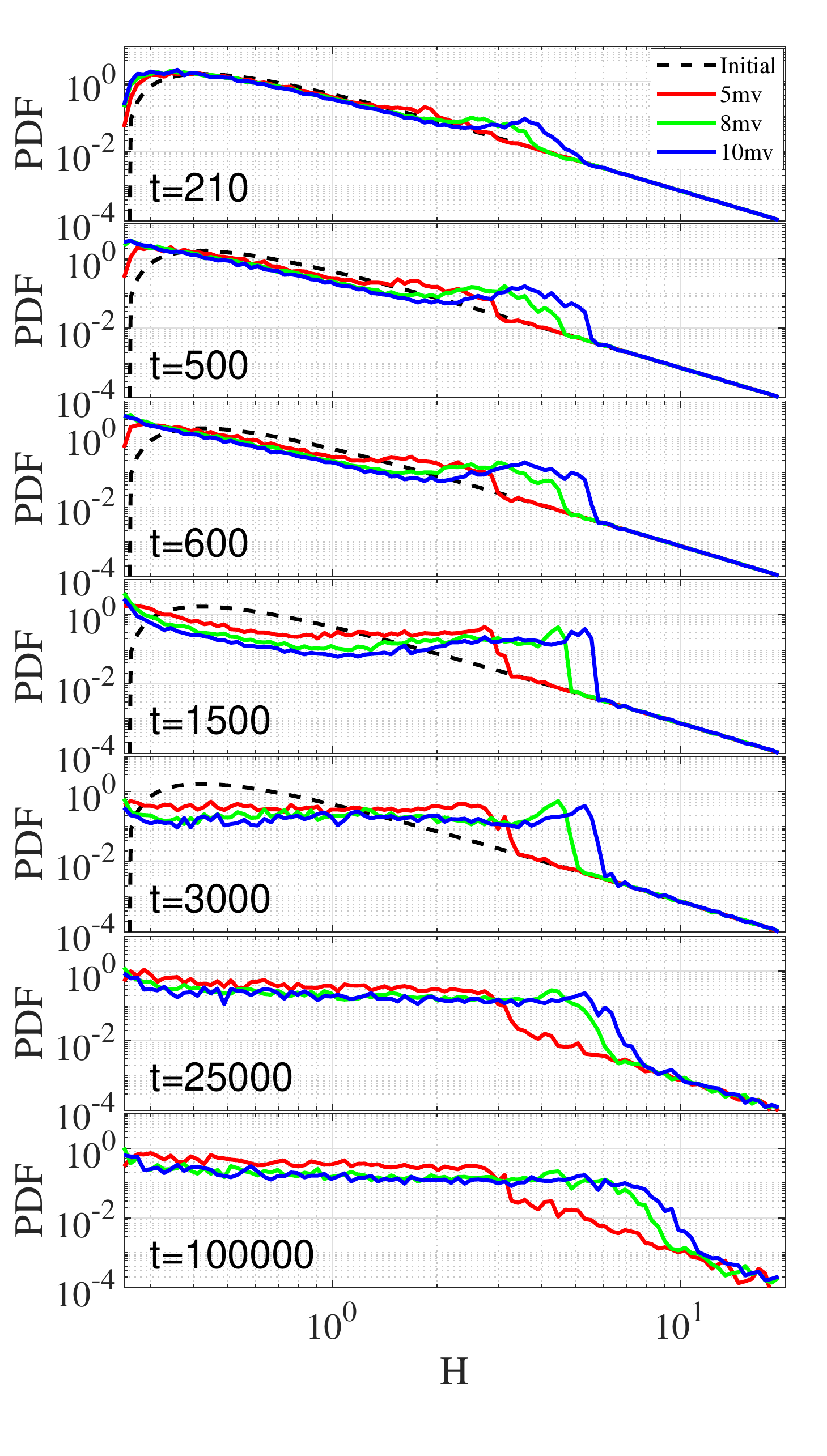}
\caption{Electron energy distribution for seven time moments for three different wave amplitudes; dashed lines show initial distribution, red, green and blue solid lines show SDE results for $E_w=5$, $E_w=8$ and $E_w=10$ respectively.}
\label{fig6}
\end{figure}

Figure \ref{fig6} compares results of SDE approach for three wave amplitudes. Larger wave electric field allows waves to keep particles trapped for a longer time and accelerate them to higher energies. This determines the energy range of the nonlinear wave-particle interaction, i.e. the energy range of electron distribution relaxation to the plateau\cite{Artemyev19:pd}. The time-scale of the this relaxation, however, does not strongly depend on the wave amplitude: to the moment $t\sim 3000$ for all three simulations we can see well formed plateau. After fast relaxation driven by nonlinear wave-particle interaction, the distribution function evolves due to diffusion on waves. The energy range of the resonant interaction, where particle diffusion occurs, is much broader than the energy range of the nonlinear wave-particle interaction. Thus, shaped by such nonlinear interactions the sharp gradient at the energies of trapped particle release starts slowly drift and relaxing to higher energies. The time scale of such diffusive relaxation strongly depends on the wave amplitude, as it should be for the quasi-linear diffusion\cite{bookSchulz&anzerotti74}.

\section{Multi-resonance system}
\begin{figure*}
\centering
\includegraphics[width=0.95\textwidth]{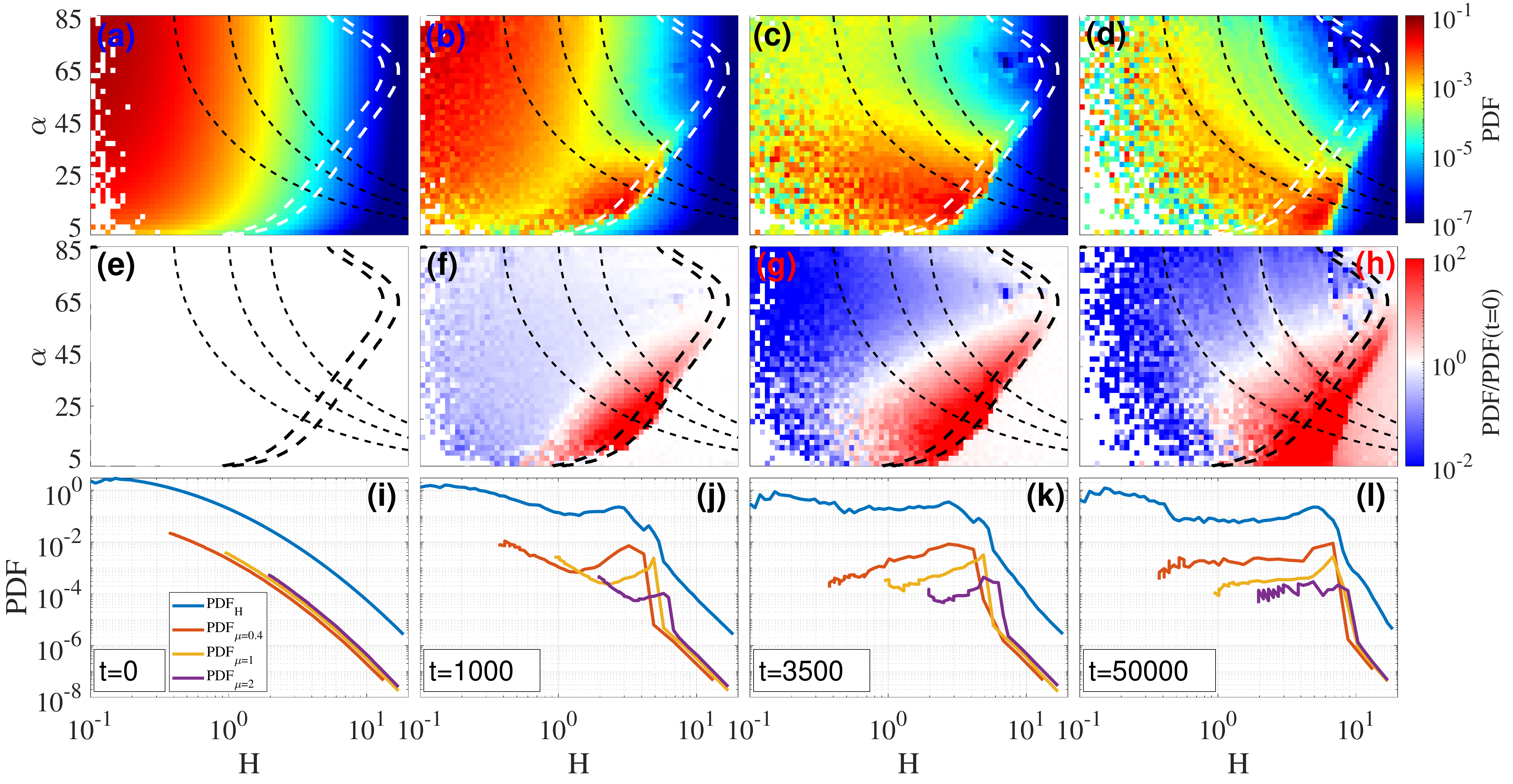}
\caption{Electron distributions in (energy, pitch-angle) space for four moments of time (SDE approach results). Panels (a)-(d) show electron distribution functions, panels (e)-(h) show ratio of PDF for fixed time to the initial one, panels (i)-(l) show four 1D profiles for PDF integrated over pitch-angles (blue line), and three PDF for the fixed magnetic moments. Thin dashed lines in panels (a)-(h) show contours of the constant magnetic moment, bold dashed lines show boundary of the nonlinear wave-particle interaction regions for two waves with $\omega/\Omega_{ce}(0)=0.05,0.55$.}
\label{fig7}
\end{figure*}

To test the proposed DSE approach we use the single wave approximation (shown in Figs.\ref{fig4}-\ref{fig6}) that allows to consider 1D distributions for constant magnetic moment. In the realistic space plasma systems, like the Earth radiation belts, particles resonate with a wave ensemble characterized by a certain range of frequencies and amplitudes. However, these interactions with different waves are rarely overlapped in time, because observed intense waves are quite coherent\cite{Cattell08, Agapitov13:jgr, Tsurutani20:jgr}. Thus, each wave-particle resonant interaction can be considered separately, without resonant destruction by other waves. In this case, the effects of multiple resonant interaction with different waves can be included into our model. To demonstrate that the proposed approach works well for such wave ensemble, we consider a typical spectrum of whistler-mode waves in the radiation belts \cite{Agapitov13:jgr,Meredith12,Li11}: wave intensity distribution $\mathcal{E}(\omega)$ is given by the Gaussian function with the peak at $\omega_m/\Omega_{ce}(0)=0.35$ and upper and lower cutoffs $\omega_{\pm}/\Omega_{ce}(0)=0.55,0.05$. Assuming that electrons interact only with one wave per quarter of the bounce period (this assumption is based on the occurrence rate of very oblique whistler-mode wave-packets, see discussion in Refs. \onlinecite{Artemyev12:pop:nondiffusion, Agapitov14:jgr:acceleration}), we represent this wave spectrum as a sum of five waves with $\omega_n=0.05\Omega_{ce}(0)\cdot n$, $n=3,5,7,9,11$. For each wave we evaluate $\Delta H$-distributions. In evaluation of electron trajectories within the SDE approach, at each iteration we use the probability distribution $\mathcal{E}(\omega)$ to determine the wave frequency and corresponding $\Delta H$-distribution. Results of such simulation are shown in Fig. \ref{fig7} for 2D electron distributions in energy/pitch-angle space (where the equatorial pitch-angle is calculated as $\alpha={\rm asin}\sqrt{2\mu/H}$).

Top line of panels (a)-(d) shows evolution of electron distribution starting with
$g(H/H_m)\sin\alpha$ ($H_m=0.25$ and $g(H)$ shown in Fig. \ref{fig7}(a)). Electrons are moving along constant magnetic moment curves (those are resonant curves for the Landau resonance\cite{bookSchulz&anzerotti74}) to higher energies due to trapping. This quickly increases the electron phase space density at low pitch-angles and higher energies (see the second line of panels (e)-(h)). Then the phase bunching (scattering) drift returns electrons to lower energies/higher pitch-angles. These two processes form the plateau in 2D space within the region of nonlinear resonances (bounded by dashed curves). Such plateau is well seen for each line of the constant magnetic moment, and for the pitch-angle averaged distribution (see the bottom line of panels (i)-(j)).


\section{Discussion}
In this study we proposed the approach for evaluation of the long-term electron distribution dynamics in systems with nonlinear resonant wave-particle interaction. The main advance of this approach is that it is not based on analytical models of such interactions, because such models generally require a significant oversimplification of wave fields (see discussion in, e.g., Refs. \onlinecite{Artemyev19:cnsns, Zheng19:emic}). However, this advantage is also associated with the increase of the potential uncertainties: any errors with numerical evaluation of initial $\Delta H$ distributions will grow in multiple applications of this distributions for stochastic trajectory tracing. Thus, the crucially important point is to verify the derived $\Delta H$ distributions with the full test particle simulations for reasonably short time intervals (see, e.g., Figs. \ref{fig4}- \ref{fig6}). Of course, such a comparison cannot totally exclude numerical uncertainties, but it should confirm that $\Delta H$ distributions describe one of the main properties of nonlinear wave-particle interaction systems: a fine balance between the probability of phase trapping and an amplitude of particle drifts due to the phase bunching \cite{Shklyar11:angeo, Artemyev16:pop:letter, Shklyar20}. This balance guarantees that all trapped and accelerated particles have a finite probability to return to their initial energies due to the phase bunching effect, i.e. the Poincaré recurrence theorem should be hold\cite{bookAKN06}. A presence of such recurrence of resonant electron motion in the phase space suggests that numerically obtained $\Delta H$ distributions provide an adequate description of the system. Thus, a good test of $\Delta H$ distributions and SDE approach for a particular system would be a long-term running of a single trajectory with the requirement that this trajectory fill the entire phase space volume of resonant interactions (see discussion in Ref. \onlinecite{Artemyev20:rcd}).

The proposed approach for simulations of long-term dynamics of electrons in the system with multiple nonlinear resonances aims to substitute a more traditional approach of solutions of diffusive (Fokker-Plank) equation with the quasi-linear diffusion rates \cite{Shprits08:JASTP_local, bookSchulz&anzerotti74}. Being new for radiation belt simulations, however, this approach with non-Gaussian statistics included into SDEs resembles well the Monte-Carlo codes modeling energetic particle acceleration on the astrophysical shock \cite{Bykov17}. Although the origin of non-Gaussian statistics of energy variations can be quite different for these two set of systems (see discussion on non-Gaussian dynamics on shocks in Refs.\onlinecite{Zimbardo&Perri13, Zimbardo&Perri20}), the numerical approaches for modeling energetic particle acceleration can be quite similar. The same approach of SDE equations with non-Gaussian statistics of energy variations can be applied to the problem of non-diffusive particle acceleration in strong turbulence of solar corona \cite{Bykov&Fleishman09, Vlahos&Isliker19}, where the most advanced models generalize Fokker-Plank equation to the fractal Fokker-Plank equation \cite{Zimbardo17, Isliker17, Isliker17:apj} to describe a finite statistics of large energy changes.

To conclude, we have considered the SDE-based approach for tracing a large particle ensemble in the system with multiple nonlinear resonant interactions. This approach generalizes the SDE approach for the quasi-linear diffusion \cite{Tao08:stochastic} by including nonlinear resonant effects with the non-Gaussian probability distribution of energy/pitch-angle changes. The approach has successfully verified with test particle simulations for the electron ensemble resonating with oblique whistler-mode waves in the Earth radiation belts. This approach can be used for simulation of the long-term evolution of electron distribution function in space plasma systems where electron dynamics are essentially determined by a combination of multiple resonant effects and large-scale adiabatic processes.

\begin{acknowledgements}
This work has been supported by the Basis Foundation grant \#19-1-5-141-1 (A.S.L).
\end{acknowledgements}

\section*{Data Availability}
This is theoretical study, and all figures are plotted using numerical solutions of equations provided with the paper. The data used for figures and findings in this study are available from the corresponding author upon reasonable request.

\bibliographystyle{elsarticle-harv}

\begin{thebibliography}{120}
\expandafter\ifx\csname natexlab\endcsname\relax\def\natexlab#1{#1}\fi
\providecommand{\url}[1]{\texttt{#1}}
\providecommand{\href}[2]{#2}
\providecommand{\path}[1]{#1}
\providecommand{\DOIprefix}{}
\providecommand{\ArXivprefix}{arXiv:}
\providecommand{\URLprefix}{URL: }
\providecommand{\Pubmedprefix}{pmid:}
\providecommand{\doi}[1]{\href{http://dx.doi.org/#1}{\path{#1}}}
\providecommand{\Pubmed}[1]{\href{pmid:#1}{\path{#1}}}
\providecommand{\bibinfo}[2]{#2}
\ifx\xfnm\relax \def\xfnm[#1]{\unskip,\space#1}\fi
\bibitem[{{Agapitov} et~al.(2013){Agapitov}, {Artemyev}, {Krasnoselskikh},
  {Khotyaintsev}, {Mourenas}, {Breuillard}, {Balikhin} and
  {Rolland}}]{Agapitov13:jgr}
\bibinfo{author}{{Agapitov}, O.V.}, \bibinfo{author}{{Artemyev}, A.},
  \bibinfo{author}{{Krasnoselskikh}, V.}, \bibinfo{author}{{Khotyaintsev},
  Y.V.}, \bibinfo{author}{{Mourenas}, D.}, \bibinfo{author}{{Breuillard}, H.},
  \bibinfo{author}{{Balikhin}, M.}, \bibinfo{author}{{Rolland}, G.},
  \bibinfo{year}{2013}.
\newblock \bibinfo{title}{{Statistics of whistler mode waves in the outer
  radiation belt: Cluster STAFF-SA measurements}}.
\newblock \bibinfo{journal}{\jgr} \bibinfo{volume}{118},
  \bibinfo{pages}{3407--3420}.
\newblock \DOIprefix\doi{10.1002/jgra.50312}.
\bibitem[{{Agapitov} et~al.(2014){Agapitov}, {Artemyev}, {Mourenas},
  {Krasnoselskikh}, {Bonnell}, {Le Contel}, {Cully} and
  {Angelopoulos}}]{Agapitov14:jgr:acceleration}
\bibinfo{author}{{Agapitov}, O.V.}, \bibinfo{author}{{Artemyev}, A.},
  \bibinfo{author}{{Mourenas}, D.}, \bibinfo{author}{{Krasnoselskikh}, V.},
  \bibinfo{author}{{Bonnell}, J.}, \bibinfo{author}{{Le Contel}, O.},
  \bibinfo{author}{{Cully}, C.M.}, \bibinfo{author}{{Angelopoulos}, V.},
  \bibinfo{year}{2014}.
\newblock \bibinfo{title}{{The quasi-electrostatic mode of chorus waves and
  electron nonlinear acceleration}}.
\newblock \bibinfo{journal}{\jgr} \bibinfo{volume}{119},
  \bibinfo{pages}{1606--1626}.
\newblock \DOIprefix\doi{10.1002/2013JA019223}.
\bibitem[{{Agapitov} et~al.(2015){Agapitov}, {Artemyev}, {Mourenas}, {Mozer}
  and {Krasnoselskikh}}]{Agapitov15:grl:acceleration}
\bibinfo{author}{{Agapitov}, O.V.}, \bibinfo{author}{{Artemyev}, A.V.},
  \bibinfo{author}{{Mourenas}, D.}, \bibinfo{author}{{Mozer}, F.S.},
  \bibinfo{author}{{Krasnoselskikh}, V.}, \bibinfo{year}{2015}.
\newblock \bibinfo{title}{{Nonlinear local parallel acceleration of electrons
  through Landau trapping by oblique whistler mode waves in the outer radiation
  belt}}.
\newblock \bibinfo{journal}{\grl} \bibinfo{volume}{42}, \bibinfo{pages}{10}.
\newblock \DOIprefix\doi{10.1002/2015GL066887}.
\bibitem[{{Agapitov} et~al.(2018){Agapitov}, {Mourenas}, {Artemyev}, {Mozer},
  {Hospodarsky}, {Bonnell} and {Krasnoselskikh}}]{Agapitov18:jgr}
\bibinfo{author}{{Agapitov}, O.V.}, \bibinfo{author}{{Mourenas}, D.},
  \bibinfo{author}{{Artemyev}, A.V.}, \bibinfo{author}{{Mozer}, F.S.},
  \bibinfo{author}{{Hospodarsky}, G.}, \bibinfo{author}{{Bonnell}, J.},
  \bibinfo{author}{{Krasnoselskikh}, V.}, \bibinfo{year}{2018}.
\newblock \bibinfo{title}{{Synthetic Empirical Chorus Wave Model From Combined
  Van Allen Probes and Cluster Statistics}}.
\newblock \bibinfo{journal}{Journal of Geophysical Research (Space Physics)}
  \bibinfo{volume}{123}, \bibinfo{pages}{297--314}.
\newblock \DOIprefix\doi{10.1002/2017JA024843}.
\bibitem[{{Albert}(1993)}]{Albert93}
\bibinfo{author}{{Albert}, J.M.}, \bibinfo{year}{1993}.
\newblock \bibinfo{title}{{Cyclotron resonance in an inhomogeneous magnetic
  field}}.
\newblock \bibinfo{journal}{Physics of Fluids B} \bibinfo{volume}{5},
  \bibinfo{pages}{2744--2750}.
\newblock \DOIprefix\doi{10.1063/1.860715}.
\bibitem[{{Albert}(2001)}]{Albert01}
\bibinfo{author}{{Albert}, J.M.}, \bibinfo{year}{2001}.
\newblock \bibinfo{title}{{Comparison of pitch angle diffusion by turbulent and
  monochromatic whistler waves}}.
\newblock \bibinfo{journal}{\jgr} \bibinfo{volume}{106},
  \bibinfo{pages}{8477--8482}.
\newblock \DOIprefix\doi{10.1029/2000JA000304}.
\bibitem[{{Albert}(2010)}]{Albert10}
\bibinfo{author}{{Albert}, J.M.}, \bibinfo{year}{2010}.
\newblock \bibinfo{title}{{Diffusion by one wave and by many waves}}.
\newblock \bibinfo{journal}{\jgr} \bibinfo{volume}{115}, \bibinfo{pages}{0}.
\newblock \DOIprefix\doi{10.1029/2009JA014732}.
\bibitem[{{Albert} et~al.(2013){Albert}, {Tao} and {Bortnik}}]{Albert13:AGU}
\bibinfo{author}{{Albert}, J.M.}, \bibinfo{author}{{Tao}, X.},
  \bibinfo{author}{{Bortnik}, J.}, \bibinfo{year}{2013}.
\newblock \bibinfo{title}{{Aspects of Nonlinear Wave-Particle Interactions}},
  in: \bibinfo{editor}{{Summers}, D.}, \bibinfo{editor}{{Mann}, I.U.},
  \bibinfo{editor}{{Baker}, D.N.}, \bibinfo{editor}{{Schulz}, M.} (Eds.),
  \bibinfo{booktitle}{Dynamics of the Earth's Radiation Belts and Inner
  Magnetosphere}.
\newblock \DOIprefix\doi{10.1029/2012GM001324}.
\bibitem[{{Allanson} et~al.(2020){Allanson}, {Watt}, {Ratcliffe}, {Allison},
  {Meredith}, {Bentley}, {Ross} and {Glauert}}]{Allanson20}
\bibinfo{author}{{Allanson}, O.}, \bibinfo{author}{{Watt}, C.E.J.},
  \bibinfo{author}{{Ratcliffe}, H.}, \bibinfo{author}{{Allison}, H.J.},
  \bibinfo{author}{{Meredith}, N.P.}, \bibinfo{author}{{Bentley}, S.N.},
  \bibinfo{author}{{Ross}, J.P.J.}, \bibinfo{author}{{Glauert}, S.A.},
  \bibinfo{year}{2020}.
\newblock \bibinfo{title}{{Particle-in-Cell Experiments Examine Electron
  Diffusion by Whistler-Mode Waves: 2. Quasi-Linear and Nonlinear Dynamics}}.
\newblock \bibinfo{journal}{Journal of Geophysical Research (Space Physics)}
  \bibinfo{volume}{125}, \bibinfo{pages}{e27949}.
\newblock \DOIprefix\doi{10.1029/2020JA027949}.
\bibitem[{{Allanson} et~al.(2019){Allanson}, {Watt}, {Ratcliffe}, {Meredith},
  {Allison}, {Bentley}, {Bloch} and {Glauert}}]{Allanson19}
\bibinfo{author}{{Allanson}, O.}, \bibinfo{author}{{Watt}, C.E.J.},
  \bibinfo{author}{{Ratcliffe}, H.}, \bibinfo{author}{{Meredith}, N.P.},
  \bibinfo{author}{{Allison}, H.J.}, \bibinfo{author}{{Bentley}, S.N.},
  \bibinfo{author}{{Bloch}, T.}, \bibinfo{author}{{Glauert}, S.A.},
  \bibinfo{year}{2019}.
\newblock \bibinfo{title}{{Particle-in-cell Experiments Examine Electron
  Diffusion by Whistler-mode Waves: 1. Benchmarking With a Cold Plasma}}.
\newblock \bibinfo{journal}{Journal of Geophysical Research (Space Physics)}
  \bibinfo{volume}{124}, \bibinfo{pages}{8893--8912}.
\newblock \DOIprefix\doi{10.1029/2019JA027088}.
\bibitem[{{Allison} and {Shprits}(2020)}]{Allison&Shprits20}
\bibinfo{author}{{Allison}, H.J.}, \bibinfo{author}{{Shprits}, Y.Y.},
  \bibinfo{year}{2020}.
\newblock \bibinfo{title}{{Local heating of radiation belt electrons to
  ultra-relativistic energies}}.
\newblock \bibinfo{journal}{Nature Communications} \bibinfo{volume}{11},
  \bibinfo{pages}{4533}.
\newblock \DOIprefix\doi{10.1038/s41467-020-18053-z}.
\bibitem[{{Andronov} and {Trakhtengerts}(1964)}]{Andronov&Trakhtengerts64}
\bibinfo{author}{{Andronov}, A.A.}, \bibinfo{author}{{Trakhtengerts}, V.Y.},
  \bibinfo{year}{1964}.
\newblock \bibinfo{title}{{Kinetic instability of the Earth's outer radiation
  belt}}.
\newblock \bibinfo{journal}{Geomagnetism and Aeronomy} \bibinfo{volume}{4},
  \bibinfo{pages}{233--242}.
\bibitem[{{Arnold} et~al.(2006){Arnold}, {Kozlov} and {Neishtadt}}]{bookAKN06}
\bibinfo{author}{{Arnold}, V.I.}, \bibinfo{author}{{Kozlov}, V.V.},
  \bibinfo{author}{{Neishtadt}, A.I.}, \bibinfo{year}{2006}.
\newblock \bibinfo{title}{Mathematical Aspects of Classical and Celestial
  Mechanics}.
\newblock Dynamical Systems III. Encyclopedia of Mathematical Sciences.
  \bibinfo{edition}{3rd} ed., \bibinfo{publisher}{Springer-Verlag},
  \bibinfo{address}{New York}.
\bibitem[{{Artemyev} et~al.(2016a){Artemyev}, {Agapitov}, {Mourenas},
  {Krasnoselskikh}, {Shastun} and {Mozer}}]{Artemyev16:ssr}
\bibinfo{author}{{Artemyev}, A.V.}, \bibinfo{author}{{Agapitov}, O.},
  \bibinfo{author}{{Mourenas}, D.}, \bibinfo{author}{{Krasnoselskikh}, V.},
  \bibinfo{author}{{Shastun}, V.}, \bibinfo{author}{{Mozer}, F.},
  \bibinfo{year}{2016}a.
\newblock \bibinfo{title}{{Oblique Whistler-Mode Waves in the Earth's Inner
  Magnetosphere: Energy Distribution, Origins, and Role in Radiation Belt
  Dynamics}}.
\newblock \bibinfo{journal}{\ssr} \bibinfo{volume}{200},
  \bibinfo{pages}{261--355}.
\newblock \DOIprefix\doi{10.1007/s11214-016-0252-5}.
\bibitem[{{Artemyev} et~al.(2012){Artemyev}, {Krasnoselskikh}, {Agapitov},
  {Mourenas} and {Rolland}}]{Artemyev12:pop:nondiffusion}
\bibinfo{author}{{Artemyev}, A.V.}, \bibinfo{author}{{Krasnoselskikh}, V.},
  \bibinfo{author}{{Agapitov}, O.}, \bibinfo{author}{{Mourenas}, D.},
  \bibinfo{author}{{Rolland}, G.}, \bibinfo{year}{2012}.
\newblock \bibinfo{title}{{Non-diffusive resonant acceleration of electrons in
  the radiation belts.}}
\newblock \bibinfo{journal}{Physics of Plasmas} \bibinfo{volume}{19},
  \bibinfo{pages}{122901}.
\newblock \DOIprefix\doi{10.1063/1.4769726}.
\bibitem[{{Artemyev} et~al.(2018a){Artemyev}, {Neishtadt}, {Vainchtein},
  {Vasiliev}, {Vasko} and {Zelenyi}}]{Artemyev18:cnsns}
\bibinfo{author}{{Artemyev}, A.V.}, \bibinfo{author}{{Neishtadt}, A.I.},
  \bibinfo{author}{{Vainchtein}, D.L.}, \bibinfo{author}{{Vasiliev}, A.A.},
  \bibinfo{author}{{Vasko}, I.Y.}, \bibinfo{author}{{Zelenyi}, L.M.},
  \bibinfo{year}{2018}a.
\newblock \bibinfo{title}{{Trapping (capture) into resonance and scattering on
  resonance: Summary of results for space plasma systems}}.
\newblock \bibinfo{journal}{Communications in Nonlinear Science and Numerical
  Simulations} \bibinfo{volume}{65}, \bibinfo{pages}{111--160}.
\newblock \DOIprefix\doi{10.1016/j.cnsns.2018.05.004}.
\bibitem[{{Artemyev} et~al.(2019a){Artemyev}, {Neishtadt} and
  {Vasiliev}}]{Artemyev19:pd}
\bibinfo{author}{{Artemyev}, A.V.}, \bibinfo{author}{{Neishtadt}, A.I.},
  \bibinfo{author}{{Vasiliev}, A.A.}, \bibinfo{year}{2019}a.
\newblock \bibinfo{title}{{Kinetic equation for nonlinear wave-particle
  interaction: Solution properties and asymptotic dynamics}}.
\newblock \bibinfo{journal}{Physica D Nonlinear Phenomena}
  \bibinfo{volume}{393}, \bibinfo{pages}{1--8}.
\newblock \DOIprefix\doi{10.1016/j.physd.2018.12.007},
  \href{http://arxiv.org/abs/1809.03743}{{\tt arXiv:1809.03743}}.
\bibitem[{{Artemyev} et~al.(2020a){Artemyev}, {Neishtadt} and
  {Vasiliev}}]{Artemyev20:rcd}
\bibinfo{author}{{Artemyev}, A.V.}, \bibinfo{author}{{Neishtadt}, A.I.},
  \bibinfo{author}{{Vasiliev}, A.A.}, \bibinfo{year}{2020}a.
\newblock \bibinfo{title}{{A Map for Systems with Resonant Trappings and
  Scatterings}}.
\newblock \bibinfo{journal}{Regular and Chaotic Dynamics} \bibinfo{volume}{25},
  \bibinfo{pages}{2--10}.
\newblock \DOIprefix\doi{10.1134/S1560354720010025}.
\bibitem[{{Artemyev} et~al.(2020b){Artemyev}, {Neishtadt} and
  {Vasiliev}}]{Artemyev20:pop}
\bibinfo{author}{{Artemyev}, A.V.}, \bibinfo{author}{{Neishtadt}, A.I.},
  \bibinfo{author}{{Vasiliev}, A.A.}, \bibinfo{year}{2020}b.
\newblock \bibinfo{title}{{Mapping for nonlinear electron interaction with
  whistler-mode waves}}.
\newblock \bibinfo{journal}{Physics of Plasmas} \bibinfo{volume}{27},
  \bibinfo{pages}{042902}.
\newblock \DOIprefix\doi{10.1063/1.5144477},
  \href{http://arxiv.org/abs/1911.11459}{{\tt arXiv:1911.11459}}.
\bibitem[{{Artemyev} et~al.(2016b){Artemyev}, {Neishtadt}, {Vasiliev} and
  {Mourenas}}]{Artemyev16:pop:letter}
\bibinfo{author}{{Artemyev}, A.V.}, \bibinfo{author}{{Neishtadt}, A.I.},
  \bibinfo{author}{{Vasiliev}, A.A.}, \bibinfo{author}{{Mourenas}, D.},
  \bibinfo{year}{2016}b.
\newblock \bibinfo{title}{{Kinetic equation for nonlinear resonant
  wave-particle interaction}}.
\newblock \bibinfo{journal}{Physics of Plasmas} \bibinfo{volume}{23},
  \bibinfo{pages}{090701}.
\newblock \DOIprefix\doi{10.1063/1.4962526}.
\bibitem[{{Artemyev} et~al.(2017){Artemyev}, {Neishtadt}, {Vasiliev} and
  {Mourenas}}]{Artemyev17:pre}
\bibinfo{author}{{Artemyev}, A.V.}, \bibinfo{author}{{Neishtadt}, A.I.},
  \bibinfo{author}{{Vasiliev}, A.A.}, \bibinfo{author}{{Mourenas}, D.},
  \bibinfo{year}{2017}.
\newblock \bibinfo{title}{{Probabilistic approach to nonlinear wave-particle
  resonant interaction}}.
\newblock \bibinfo{journal}{\pre} \bibinfo{volume}{95},
  \bibinfo{pages}{023204}.
\newblock \DOIprefix\doi{10.1103/PhysRevE.95.023204}.
\bibitem[{{Artemyev} et~al.(2018b){Artemyev}, {Neishtadt}, {Vasiliev} and
  {Mourenas}}]{Artemyev18:jpp}
\bibinfo{author}{{Artemyev}, A.V.}, \bibinfo{author}{{Neishtadt}, A.I.},
  \bibinfo{author}{{Vasiliev}, A.A.}, \bibinfo{author}{{Mourenas}, D.},
  \bibinfo{year}{2018}b.
\newblock \bibinfo{title}{{Long-term evolution of electron distribution
  function due to nonlinear resonant interaction with whistler mode waves}}.
\newblock \bibinfo{journal}{Journal of Plasma Physics} \bibinfo{volume}{84},
  \bibinfo{pages}{905840206}.
\newblock \DOIprefix\doi{10.1017/S0022377818000260}.
\bibitem[{{Artemyev} et~al.(2020c){Artemyev}, {Neishtadt}, {Vasiliev}, {Zhang},
  {Mourenas} and {Vainchtein}}]{Artemyev20:jpp:arxiv}
\bibinfo{author}{{Artemyev}, A.V.}, \bibinfo{author}{{Neishtadt}, A.I.},
  \bibinfo{author}{{Vasiliev}, A.A.}, \bibinfo{author}{{Zhang}, X.J.},
  \bibinfo{author}{{Mourenas}, D.}, \bibinfo{author}{{Vainchtein}, D.},
  \bibinfo{year}{2020}c.
\newblock \bibinfo{title}{{Long-term dynamics driven by resonant wave-particle
  interactions: from Hamiltonian resonance theory to phase space mapping}}.
\newblock \bibinfo{journal}{arXiv e-prints} ,
  \bibinfo{pages}{arXiv:2011.00208}\href{http://arxiv.org/abs/2011.00208}{{\tt
  arXiv:2011.00208}}.
\bibitem[{{Artemyev} et~al.(2014a){Artemyev}, {Vasiliev}, {Mourenas},
  {Agapitov}, {Krasnoselskikh}, {Boscher} and
  {Rolland}}]{Artemyev14:grl:fast_transport}
\bibinfo{author}{{Artemyev}, A.V.}, \bibinfo{author}{{Vasiliev}, A.A.},
  \bibinfo{author}{{Mourenas}, D.}, \bibinfo{author}{{Agapitov}, O.},
  \bibinfo{author}{{Krasnoselskikh}, V.}, \bibinfo{author}{{Boscher}, D.},
  \bibinfo{author}{{Rolland}, G.}, \bibinfo{year}{2014}a.
\newblock \bibinfo{title}{{Fast transport of resonant electrons in phase space
  due to nonlinear trapping by whistler waves}}.
\newblock \bibinfo{journal}{\grl} \bibinfo{volume}{41},
  \bibinfo{pages}{5727--5733}.
\newblock \DOIprefix\doi{10.1002/2014GL061380}.
\bibitem[{{Artemyev} et~al.(2014b){Artemyev}, {Vasiliev}, {Mourenas},
  {Agapitov} and {Krasnoselskikh}}]{Artemyev14:pop}
\bibinfo{author}{{Artemyev}, A.V.}, \bibinfo{author}{{Vasiliev}, A.A.},
  \bibinfo{author}{{Mourenas}, D.}, \bibinfo{author}{{Agapitov}, O.V.},
  \bibinfo{author}{{Krasnoselskikh}, V.V.}, \bibinfo{year}{2014}b.
\newblock \bibinfo{title}{{Electron scattering and nonlinear trapping by
  oblique whistler waves: The critical wave intensity for nonlinear effects}}.
\newblock \bibinfo{journal}{Physics of Plasmas} \bibinfo{volume}{21},
  \bibinfo{pages}{102903}.
\newblock \DOIprefix\doi{10.1063/1.4897945}.
\bibitem[{{Artemyev} et~al.(2019b){Artemyev}, {Vasiliev} and
  {Neishtadt}}]{Artemyev19:cnsns}
\bibinfo{author}{{Artemyev}, A.V.}, \bibinfo{author}{{Vasiliev}, A.A.},
  \bibinfo{author}{{Neishtadt}, A.I.}, \bibinfo{year}{2019}b.
\newblock \bibinfo{title}{{Charged particle nonlinear resonance with localized
  electrostatic wave-packets}}.
\newblock \bibinfo{journal}{Communications in Nonlinear Science and Numerical
  Simulations} \bibinfo{volume}{72}, \bibinfo{pages}{392--406}.
\newblock \DOIprefix\doi{10.1016/j.cnsns.2019.01.011}.
\bibitem[{{Ashour-Abdalla} et~al.(2013){Ashour-Abdalla}, {Schriver}, {Alaoui},
  {Richard}, {Walker}, {Goldstein}, {Donovan} and {Zhou}}]{Maha13:lsk}
\bibinfo{author}{{Ashour-Abdalla}, M.}, \bibinfo{author}{{Schriver}, D.},
  \bibinfo{author}{{Alaoui}, M.E.}, \bibinfo{author}{{Richard}, R.},
  \bibinfo{author}{{Walker}, R.}, \bibinfo{author}{{Goldstein}, M.L.},
  \bibinfo{author}{{Donovan}, E.}, \bibinfo{author}{{Zhou}, M.},
  \bibinfo{year}{2013}.
\newblock \bibinfo{title}{{Direct auroral precipitation from the magnetotail
  during substorms}}.
\newblock \bibinfo{journal}{\grl} \bibinfo{volume}{40},
  \bibinfo{pages}{3787--3792}.
\newblock \DOIprefix\doi{10.1002/grl.50635}.
\bibitem[{{Balikhin} et~al.(1993){Balikhin}, {Gedalin} and
  {Petrukovich}}]{Balikhin93}
\bibinfo{author}{{Balikhin}, M.}, \bibinfo{author}{{Gedalin}, M.},
  \bibinfo{author}{{Petrukovich}, A.}, \bibinfo{year}{1993}.
\newblock \bibinfo{title}{{New mechanism for electron heating in shocks}}.
\newblock \bibinfo{journal}{Physical Review Letters} \bibinfo{volume}{70},
  \bibinfo{pages}{1259--1262}.
\newblock \DOIprefix\doi{10.1103/PhysRevLett.70.1259}.
\bibitem[{{Birn} et~al.(2014){Birn}, {Runov} and {Hesse}}]{Birn14}
\bibinfo{author}{{Birn}, J.}, \bibinfo{author}{{Runov}, A.},
  \bibinfo{author}{{Hesse}, M.}, \bibinfo{year}{2014}.
\newblock \bibinfo{title}{{Energetic electrons in dipolarization events:
  Spatial properties and anisotropy}}.
\newblock \bibinfo{journal}{Journal of Geophysical Research (Space Physics)}
  \bibinfo{volume}{119}, \bibinfo{pages}{3604--3616}.
\newblock \DOIprefix\doi{10.1002/2013JA019738}.
\bibitem[{{Breuillard} et~al.(2016){Breuillard}, {Le Contel}, {Retino},
  {Chasapis}, {Chust}, {Mirioni}, {Graham}, {Wilder}, {Cohen}, {Vaivads},
  {Khotyaintsev}, {Lindqvist}, {Marklund}, {Burch}, {Torbert}, {Ergun},
  {Goodrich}, {Macri}, {Needell}, {Chutter}, {Rau}, {Dors}, {Russell},
  {Magnes}, {Strangeway}, {Bromund}, {Plaschke}, {Fischer}, {Leinweber},
  {Anderson}, {Le}, {Slavin}, {Kepko}, {Baumjohann}, {Mauk}, {Fuselier} and
  {Nakamura}}]{Breuillard16}
\bibinfo{author}{{Breuillard}, H.}, \bibinfo{author}{{Le Contel}, O.},
  \bibinfo{author}{{Retino}, A.}, \bibinfo{author}{{Chasapis}, A.},
  \bibinfo{author}{{Chust}, T.}, \bibinfo{author}{{Mirioni}, L.},
  \bibinfo{author}{{Graham}, D.B.}, \bibinfo{author}{{Wilder}, F.D.},
  \bibinfo{author}{{Cohen}, I.}, \bibinfo{author}{{Vaivads}, A.},
  \bibinfo{author}{{Khotyaintsev}, Y.V.}, \bibinfo{author}{{Lindqvist}, P.A.},
  \bibinfo{author}{{Marklund}, G.T.}, \bibinfo{author}{{Burch}, J.L.},
  \bibinfo{author}{{Torbert}, R.B.}, \bibinfo{author}{{Ergun}, R.E.},
  \bibinfo{author}{{Goodrich}, K.A.}, \bibinfo{author}{{Macri}, J.},
  \bibinfo{author}{{Needell}, J.}, \bibinfo{author}{{Chutter}, M.},
  \bibinfo{author}{{Rau}, D.}, \bibinfo{author}{{Dors}, I.},
  \bibinfo{author}{{Russell}, C.T.}, \bibinfo{author}{{Magnes}, W.},
  \bibinfo{author}{{Strangeway}, R.J.}, \bibinfo{author}{{Bromund}, K.R.},
  \bibinfo{author}{{Plaschke}, F.}, \bibinfo{author}{{Fischer}, D.},
  \bibinfo{author}{{Leinweber}, H.K.}, \bibinfo{author}{{Anderson}, B.J.},
  \bibinfo{author}{{Le}, G.}, \bibinfo{author}{{Slavin}, J.A.},
  \bibinfo{author}{{Kepko}, E.L.}, \bibinfo{author}{{Baumjohann}, W.},
  \bibinfo{author}{{Mauk}, B.}, \bibinfo{author}{{Fuselier}, S.A.},
  \bibinfo{author}{{Nakamura}, R.}, \bibinfo{year}{2016}.
\newblock \bibinfo{title}{{Multispacecraft analysis of dipolarization fronts
  and associated whistler wave emissions using MMS data}}.
\newblock \bibinfo{journal}{\grl} \bibinfo{volume}{43},
  \bibinfo{pages}{7279--7286}.
\newblock \DOIprefix\doi{10.1002/2016GL069188}.
\bibitem[{{Bykov} et~al.(2017){Bykov}, {Ellison} and {Osipov}}]{Bykov17}
\bibinfo{author}{{Bykov}, A.M.}, \bibinfo{author}{{Ellison}, D.C.},
  \bibinfo{author}{{Osipov}, S.M.}, \bibinfo{year}{2017}.
\newblock \bibinfo{title}{{Nonlinear Monte Carlo model of superdiffusive shock
  acceleration with magnetic field amplification}}.
\newblock \bibinfo{journal}{\pre} \bibinfo{volume}{95},
  \bibinfo{pages}{033207}.
\newblock \DOIprefix\doi{10.1103/PhysRevE.95.033207},
  \href{http://arxiv.org/abs/1703.01160}{{\tt arXiv:1703.01160}}.
\bibitem[{{Bykov} and {Fleishman}(2009)}]{Bykov&Fleishman09}
\bibinfo{author}{{Bykov}, A.M.}, \bibinfo{author}{{Fleishman}, G.D.},
  \bibinfo{year}{2009}.
\newblock \bibinfo{title}{{Particle Acceleration by Strong Turbulence in Solar
  Flares: Theory of Spectrum Evolution}}.
\newblock \bibinfo{journal}{\apjl} \bibinfo{volume}{692},
  \bibinfo{pages}{L45--L49}.
\newblock \DOIprefix\doi{10.1088/0004-637X/692/1/L45},
  \href{http://arxiv.org/abs/0901.2677}{{\tt arXiv:0901.2677}}.
\bibitem[{{Bykov} and {Toptygin}(1993)}]{Bykov&Toptygin93}
\bibinfo{author}{{Bykov}, A.M.}, \bibinfo{author}{{Toptygin}, I.},
  \bibinfo{year}{1993}.
\newblock \bibinfo{title}{{REVIEWS OF TOPICAL PROBLEMS: Particle kinetics in
  highly turbulent plasmas (renormalization and self-consistent field
  methods)}}.
\newblock \bibinfo{journal}{Physics Uspekhi} \bibinfo{volume}{36},
  \bibinfo{pages}{1020--1052}.
\newblock \DOIprefix\doi{10.1070/PU1993v036n11ABEH002179}.
\bibitem[{{Camporeale} and {Zimbardo}(2015)}]{Camporeale&Zimbardo15}
\bibinfo{author}{{Camporeale}, E.}, \bibinfo{author}{{Zimbardo}, G.},
  \bibinfo{year}{2015}.
\newblock \bibinfo{title}{{Wave-particle interactions with parallel whistler
  waves: Nonlinear and time-dependent effects revealed by particle-in-cell
  simulations}}.
\newblock \bibinfo{journal}{Physics of Plasmas} \bibinfo{volume}{22},
  \bibinfo{pages}{092104}.
\newblock \DOIprefix\doi{10.1063/1.4929853},
  \href{http://arxiv.org/abs/1412.3229}{{\tt arXiv:1412.3229}}.
\bibitem[{{Cattell} et~al.(2008){Cattell}, {Wygant}, {Goetz}, {Kersten},
  {Kellogg}, {von Rosenvinge}, {Bale}, {Roth}, {Temerin}, {Hudson}, {Mewaldt},
  {Wiedenbeck}, {Maksimovic}, {Ergun}, {Acuna} and {Russell}}]{Cattell08}
\bibinfo{author}{{Cattell}, C.}, \bibinfo{author}{{Wygant}, J.R.},
  \bibinfo{author}{{Goetz}, K.}, \bibinfo{author}{{Kersten}, K.},
  \bibinfo{author}{{Kellogg}, P.J.}, \bibinfo{author}{{von Rosenvinge}, T.},
  \bibinfo{author}{{Bale}, S.D.}, \bibinfo{author}{{Roth}, I.},
  \bibinfo{author}{{Temerin}, M.}, \bibinfo{author}{{Hudson}, M.K.},
  \bibinfo{author}{{Mewaldt}, R.A.}, \bibinfo{author}{{Wiedenbeck}, M.},
  \bibinfo{author}{{Maksimovic}, M.}, \bibinfo{author}{{Ergun}, R.},
  \bibinfo{author}{{Acuna}, M.}, \bibinfo{author}{{Russell}, C.T.},
  \bibinfo{year}{2008}.
\newblock \bibinfo{title}{{Discovery of very large amplitude whistler-mode
  waves in Earth's radiation belts}}.
\newblock \bibinfo{journal}{\grl} \bibinfo{volume}{35}, \bibinfo{pages}{1105}.
\newblock \DOIprefix\doi{10.1029/2007GL032009}.
\bibitem[{{Cattell} et~al.(2015){Cattell}, {Breneman}, {Thaller}, {Wygant},
  {Kletzing} and {Kurth}}]{Cattell15}
\bibinfo{author}{{Cattell}, C.A.}, \bibinfo{author}{{Breneman}, A.W.},
  \bibinfo{author}{{Thaller}, S.A.}, \bibinfo{author}{{Wygant}, J.R.},
  \bibinfo{author}{{Kletzing}, C.A.}, \bibinfo{author}{{Kurth}, W.S.},
  \bibinfo{year}{2015}.
\newblock \bibinfo{title}{{Van Allen Probes observations of unusually low
  frequency whistler mode waves observed in association with moderate magnetic
  storms: Statistical study}}.
\newblock \bibinfo{journal}{\grl} \bibinfo{volume}{42},
  \bibinfo{pages}{7273--7281}.
\newblock \DOIprefix\doi{10.1002/2015GL065565}.
\bibitem[{{Demekhov} et~al.(2009){Demekhov}, {Trakhtengerts}, {Rycroft} and
  {Nunn}}]{Demekhov09}
\bibinfo{author}{{Demekhov}, A.G.}, \bibinfo{author}{{Trakhtengerts}, V.Y.},
  \bibinfo{author}{{Rycroft}, M.}, \bibinfo{author}{{Nunn}, D.},
  \bibinfo{year}{2009}.
\newblock \bibinfo{title}{{Efficiency of electron acceleration in the Earth's
  magnetosphere by whistler mode waves}}.
\newblock \bibinfo{journal}{Geomagnetism and Aeronomy} \bibinfo{volume}{49},
  \bibinfo{pages}{24--29}.
\newblock \DOIprefix\doi{10.1134/S0016793209010034}.
\bibitem[{{Demekhov} et~al.(2006){Demekhov}, {Trakhtengerts}, {Rycroft} and
  {Nunn}}]{Demekhov06}
\bibinfo{author}{{Demekhov}, A.G.}, \bibinfo{author}{{Trakhtengerts}, V.Y.},
  \bibinfo{author}{{Rycroft}, M.J.}, \bibinfo{author}{{Nunn}, D.},
  \bibinfo{year}{2006}.
\newblock \bibinfo{title}{{Electron acceleration in the magnetosphere by
  whistler-mode waves of varying frequency}}.
\newblock \bibinfo{journal}{Geomagnetism and Aeronomy} \bibinfo{volume}{46},
  \bibinfo{pages}{711--716}.
\newblock \DOIprefix\doi{10.1134/S0016793206060053}.
\bibitem[{{Drummond} and {Pines}(1962)}]{Drummond&Pines62}
\bibinfo{author}{{Drummond}, W.E.}, \bibinfo{author}{{Pines}, D.},
  \bibinfo{year}{1962}.
\newblock \bibinfo{title}{{Nonlinear stability of plasma oscillations}}.
\newblock \bibinfo{journal}{Nuclear Fusion Suppl.} \bibinfo{volume}{3},
  \bibinfo{pages}{1049--1058}.
\bibitem[{{Elkington} et~al.(2019){Elkington}, {Chan}, {Jaynes}, {Malaspina}
  and {Albert}}]{Elkington19:agu}
\bibinfo{author}{{Elkington}, S.R.}, \bibinfo{author}{{Chan}, A.A.},
  \bibinfo{author}{{Jaynes}, A.N.}, \bibinfo{author}{{Malaspina}, D.},
  \bibinfo{author}{{Albert}, J.}, \bibinfo{year}{2019}.
\newblock \bibinfo{title}{{K2: Towards a Comprehensive Simulation Framework of
  the Van Allen Radiation Belts}}, in: \bibinfo{booktitle}{AGU Fall Meeting
  Abstracts}, pp. \bibinfo{pages}{SM44B--01}.
\bibitem[{{Elkington} et~al.(2018){Elkington}, {Chan}, {Li}, {Hudson}, {Jaynes}
  and {Baker}}]{Elkington18:agu}
\bibinfo{author}{{Elkington}, S.R.}, \bibinfo{author}{{Chan}, A.A.},
  \bibinfo{author}{{Li}, Z.}, \bibinfo{author}{{Hudson}, M.K.},
  \bibinfo{author}{{Jaynes}, A.N.}, \bibinfo{author}{{Baker}, D.N.},
  \bibinfo{year}{2018}.
\newblock \bibinfo{title}{{Generalizing Global Simulations of the Radiation
  Belts: Addressing Advective and Diffusive Processes in a Common Simulation
  Framework}}, in: \bibinfo{booktitle}{AGU Fall Meeting Abstracts}, pp.
  \bibinfo{pages}{SM11B--02}.
\bibitem[{{Elkington} et~al.(2004){Elkington}, {Wiltberger}, {Chan} and
  {Baker}}]{Elkington04}
\bibinfo{author}{{Elkington}, S.R.}, \bibinfo{author}{{Wiltberger}, M.},
  \bibinfo{author}{{Chan}, A.A.}, \bibinfo{author}{{Baker}, D.N.},
  \bibinfo{year}{2004}.
\newblock \bibinfo{title}{{Physical models of the geospace radiation
  environment}}.
\newblock \bibinfo{journal}{Journal of Atmospheric and Solar-Terrestrial
  Physics} \bibinfo{volume}{66}, \bibinfo{pages}{1371--1387}.
\newblock \DOIprefix\doi{10.1016/j.jastp.2004.03.023}.
\bibitem[{{Eshetu} et~al.(2018){Eshetu}, {Lyon}, {Hudson} and
  {Wiltberger}}]{Eshetu18}
\bibinfo{author}{{Eshetu}, W.W.}, \bibinfo{author}{{Lyon}, J.G.},
  \bibinfo{author}{{Hudson}, M.K.}, \bibinfo{author}{{Wiltberger}, M.J.},
  \bibinfo{year}{2018}.
\newblock \bibinfo{title}{{Pitch Angle Scattering of Energetic Electrons by
  BBFs}}.
\newblock \bibinfo{journal}{Journal of Geophysical Research (Space Physics)}
  \bibinfo{volume}{123}, \bibinfo{pages}{9265--9274}.
\newblock \DOIprefix\doi{10.1029/2018JA025788}.
\bibitem[{{Furuya} et~al.(2008){Furuya}, {Omura} and {Summers}}]{Furuya08}
\bibinfo{author}{{Furuya}, N.}, \bibinfo{author}{{Omura}, Y.},
  \bibinfo{author}{{Summers}, D.}, \bibinfo{year}{2008}.
\newblock \bibinfo{title}{{Relativistic turning acceleration of radiation belt
  electrons by whistler mode chorus}}.
\newblock \bibinfo{journal}{\jgr} \bibinfo{volume}{113}, \bibinfo{pages}{4224}.
\newblock \DOIprefix\doi{10.1029/2007JA012478}.
\bibitem[{{Gabrielse} et~al.(2014){Gabrielse}, {Angelopoulos}, {Runov} and
  {Turner}}]{Gabrielse14}
\bibinfo{author}{{Gabrielse}, C.}, \bibinfo{author}{{Angelopoulos}, V.},
  \bibinfo{author}{{Runov}, A.}, \bibinfo{author}{{Turner}, D.L.},
  \bibinfo{year}{2014}.
\newblock \bibinfo{title}{{Statistical characteristics of particle injections
  throughout the equatorial magnetotail}}.
\newblock \bibinfo{journal}{\jgr} \bibinfo{volume}{119},
  \bibinfo{pages}{2512--2535}.
\newblock \DOIprefix\doi{10.1002/2013JA019638}.
\bibitem[{{Gabrielse} et~al.(2019){Gabrielse}, {Spanswick}, {Artemyev},
  {Nishimura}, {Runov}, {Lyons}, {Angelopoulos}, {Turner}, {Reeves},
  {McPherron} and {Donovan}}]{Gabrielse19}
\bibinfo{author}{{Gabrielse}, C.}, \bibinfo{author}{{Spanswick}, E.},
  \bibinfo{author}{{Artemyev}, A.}, \bibinfo{author}{{Nishimura}, Y.},
  \bibinfo{author}{{Runov}, A.}, \bibinfo{author}{{Lyons}, L.},
  \bibinfo{author}{{Angelopoulos}, V.}, \bibinfo{author}{{Turner}, D.L.},
  \bibinfo{author}{{Reeves}, G.D.}, \bibinfo{author}{{McPherron}, R.},
  \bibinfo{author}{{Donovan}, E.}, \bibinfo{year}{2019}.
\newblock \bibinfo{title}{{Utilizing the Heliophysics/Geospace System
  Observatory to Understand Particle Injections: Their Scale Sizes and
  Propagation Directions}}.
\newblock \bibinfo{journal}{Journal of Geophysical Research (Space Physics)}
  \bibinfo{volume}{124}, \bibinfo{pages}{5584--5609}.
\newblock \DOIprefix\doi{10.1029/2018JA025588}.
\bibitem[{{Gan} et~al.(2020a){Gan}, {Li}, {Ma}, {Albert}, {Artemyev} and
  {Bortnik}}]{Gan20:grl}
\bibinfo{author}{{Gan}, L.}, \bibinfo{author}{{Li}, W.}, \bibinfo{author}{{Ma},
  Q.}, \bibinfo{author}{{Albert}, J.M.}, \bibinfo{author}{{Artemyev}, A.V.},
  \bibinfo{author}{{Bortnik}, J.}, \bibinfo{year}{2020}a.
\newblock \bibinfo{title}{{Nonlinear Interactions Between Radiation Belt
  Electrons and Chorus Waves: Dependence on Wave Amplitude Modulation}}.
\newblock \bibinfo{journal}{\grl} \bibinfo{volume}{47},
  \bibinfo{pages}{e85987}.
\newblock \DOIprefix\doi{10.1029/2019GL085987}.
\bibitem[{{Gan} et~al.(2020b){Gan}, {Li}, {Ma}, {Artemyev} and
  {Albert}}]{Gan20:grl:II}
\bibinfo{author}{{Gan}, L.}, \bibinfo{author}{{Li}, W.}, \bibinfo{author}{{Ma},
  Q.}, \bibinfo{author}{{Artemyev}, A.V.}, \bibinfo{author}{{Albert}, J.M.},
  \bibinfo{year}{2020}b.
\newblock \bibinfo{title}{{Unraveling the Formation Mechanism for the Bursts of
  Electron Butterfly Distributions: Test Particle and Quasilinear
  Simulations}}.
\newblock \bibinfo{journal}{\grl} \bibinfo{volume}{47},
  \bibinfo{pages}{e90749}.
\newblock \DOIprefix\doi{10.1029/2020GL090749}.
\bibitem[{{Gedalin}(2020)}]{Gedalin20}
\bibinfo{author}{{Gedalin}, M.}, \bibinfo{year}{2020}.
\newblock \bibinfo{title}{{Large-scale versus Small-scale Fields in the Shock
  Front: Effect on the Particle Motion}}.
\newblock \bibinfo{journal}{\apj} \bibinfo{volume}{895}, \bibinfo{pages}{59}.
\newblock \DOIprefix\doi{10.3847/1538-4357/ab8af0}.
\bibitem[{{Horne} et~al.(2005){Horne}, {Thorne}, {Shprits}, {Meredith},
  {Glauert}, {Smith}, {Kanekal}, {Baker}, {Engebretson}, {Posch}, {Spasojevic},
  {Inan}, {Pickett} and {Decreau}}]{Horne05Nature}
\bibinfo{author}{{Horne}, R.B.}, \bibinfo{author}{{Thorne}, R.M.},
  \bibinfo{author}{{Shprits}, Y.Y.}, \bibinfo{author}{{Meredith}, N.P.},
  \bibinfo{author}{{Glauert}, S.A.}, \bibinfo{author}{{Smith}, A.J.},
  \bibinfo{author}{{Kanekal}, S.G.}, \bibinfo{author}{{Baker}, D.N.},
  \bibinfo{author}{{Engebretson}, M.J.}, \bibinfo{author}{{Posch}, J.L.},
  \bibinfo{author}{{Spasojevic}, M.}, \bibinfo{author}{{Inan}, U.S.},
  \bibinfo{author}{{Pickett}, J.S.}, \bibinfo{author}{{Decreau}, P.M.E.},
  \bibinfo{year}{2005}.
\newblock \bibinfo{title}{{Wave acceleration of electrons in the Van Allen
  radiation belts}}.
\newblock \bibinfo{journal}{Nature} \bibinfo{volume}{437},
  \bibinfo{pages}{227--230}.
\newblock \DOIprefix\doi{10.1038/nature03939}.
\bibitem[{{Hsieh} and {Omura}(2017a)}]{Hsieh&Omura17}
\bibinfo{author}{{Hsieh}, Y.K.}, \bibinfo{author}{{Omura}, Y.},
  \bibinfo{year}{2017}a.
\newblock \bibinfo{title}{{Nonlinear dynamics of electrons interacting with
  oblique whistler mode chorus in the magnetosphere}}.
\newblock \bibinfo{journal}{\jgr} \bibinfo{volume}{122},
  \bibinfo{pages}{675--694}.
\newblock \DOIprefix\doi{10.1002/2016JA023255}.
\bibitem[{{Hsieh} and {Omura}(2017b)}]{Hsieh&Omura17:radio_science}
\bibinfo{author}{{Hsieh}, Y.K.}, \bibinfo{author}{{Omura}, Y.},
  \bibinfo{year}{2017}b.
\newblock \bibinfo{title}{Study of wave-particle interactions for whistler mode
  waves at oblique angles by utilizing the gyroaveraging method}.
\newblock \bibinfo{journal}{Radio Science} \bibinfo{volume}{52},
  \bibinfo{pages}{1268--1281}.
\newblock \URLprefix \url{http://dx.doi.org/10.1002/2017RS006245},
  \DOIprefix\doi{10.1002/2017RS006245}. \bibinfo{note}{2017RS006245}.
\bibitem[{{Hudson} et~al.(2012){Hudson}, {Brito}, {Elkington}, {Kress}, {Li}
  and {Wiltberger}}]{Hudson12:simulation}
\bibinfo{author}{{Hudson}, M.}, \bibinfo{author}{{Brito}, T.},
  \bibinfo{author}{{Elkington}, S.}, \bibinfo{author}{{Kress}, B.},
  \bibinfo{author}{{Li}, Z.}, \bibinfo{author}{{Wiltberger}, M.},
  \bibinfo{year}{2012}.
\newblock \bibinfo{title}{{Radiation belt 2D and 3D simulations for CIR-driven
  storms during Carrington Rotation 2068}}.
\newblock \bibinfo{journal}{Journal of Atmospheric and Solar-Terrestrial
  Physics} \bibinfo{volume}{83}, \bibinfo{pages}{51--62}.
\newblock \DOIprefix\doi{10.1016/j.jastp.2012.03.017}.
\bibitem[{{Hudson} et~al.(2015){Hudson}, {Paral}, {Kress}, {Wiltberger},
  {Baker}, {Foster}, {Turner} and {Wygant}}]{Hudson15}
\bibinfo{author}{{Hudson}, M.K.}, \bibinfo{author}{{Paral}, J.},
  \bibinfo{author}{{Kress}, B.T.}, \bibinfo{author}{{Wiltberger}, M.},
  \bibinfo{author}{{Baker}, D.N.}, \bibinfo{author}{{Foster}, J.C.},
  \bibinfo{author}{{Turner}, D.L.}, \bibinfo{author}{{Wygant}, J.R.},
  \bibinfo{year}{2015}.
\newblock \bibinfo{title}{{Modeling CME-shock-driven storms in 2012-2013: MHD
  test particle simulations}}.
\newblock \bibinfo{journal}{\jgr} \bibinfo{volume}{120},
  \bibinfo{pages}{1168--1181}.
\newblock \DOIprefix\doi{10.1002/2014JA020833}.
\bibitem[{{Hull} et~al.(2012){Hull}, {Muschietti}, {Oka}, {Larson}, {Mozer},
  {Chaston}, {Bonnell} and {Hospodarsky}}]{Hull12}
\bibinfo{author}{{Hull}, A.J.}, \bibinfo{author}{{Muschietti}, L.},
  \bibinfo{author}{{Oka}, M.}, \bibinfo{author}{{Larson}, D.E.},
  \bibinfo{author}{{Mozer}, F.S.}, \bibinfo{author}{{Chaston}, C.C.},
  \bibinfo{author}{{Bonnell}, J.W.}, \bibinfo{author}{{Hospodarsky}, G.B.},
  \bibinfo{year}{2012}.
\newblock \bibinfo{title}{{Multiscale whistler waves within Earth's
  perpendicular bow shock}}.
\newblock \bibinfo{journal}{\jgr} \bibinfo{volume}{117},
  \bibinfo{pages}{12104}.
\newblock \DOIprefix\doi{10.1029/2012JA017870}.
\bibitem[{{Isliker} et~al.(2017a){Isliker}, {Pisokas}, {Vlahos} and
  {Anastasiadis}}]{Isliker17:apj}
\bibinfo{author}{{Isliker}, H.}, \bibinfo{author}{{Pisokas}, T.},
  \bibinfo{author}{{Vlahos}, L.}, \bibinfo{author}{{Anastasiadis}, A.},
  \bibinfo{year}{2017}a.
\newblock \bibinfo{title}{{Particle Acceleration and Fractional Transport in
  Turbulent Reconnection}}.
\newblock \bibinfo{journal}{\apj} \bibinfo{volume}{849}, \bibinfo{pages}{35}.
\newblock \DOIprefix\doi{10.3847/1538-4357/aa8ee8},
  \href{http://arxiv.org/abs/1709.08269}{{\tt arXiv:1709.08269}}.
\bibitem[{{Isliker} et~al.(2017b){Isliker}, {Vlahos} and
  {Constantinescu}}]{Isliker17}
\bibinfo{author}{{Isliker}, H.}, \bibinfo{author}{{Vlahos}, L.},
  \bibinfo{author}{{Constantinescu}, D.}, \bibinfo{year}{2017}b.
\newblock \bibinfo{title}{{Fractional Transport in Strongly Turbulent
  Plasmas}}.
\newblock \bibinfo{journal}{\prl} \bibinfo{volume}{119},
  \bibinfo{pages}{045101}.
\newblock \DOIprefix\doi{10.1103/PhysRevLett.119.045101},
  \href{http://arxiv.org/abs/1707.01526}{{\tt arXiv:1707.01526}}.
\bibitem[{{Itin} et~al.(2000){Itin}, {Neishtadt} and {Vasiliev}}]{Itin00}
\bibinfo{author}{{Itin}, A.P.}, \bibinfo{author}{{Neishtadt}, A.I.},
  \bibinfo{author}{{Vasiliev}, A.A.}, \bibinfo{year}{2000}.
\newblock \bibinfo{title}{{Captures into resonance and scattering on resonance
  in dynamics of a charged relativistic particle in magnetic field and
  electrostatic wave}}.
\newblock \bibinfo{journal}{Physica D: Nonlinear Phenomena}
  \bibinfo{volume}{141}, \bibinfo{pages}{281--296}.
\newblock \DOIprefix\doi{10.1016/S0167-2789(00)00039-7}.
\bibitem[{{Karimabadi} et~al.(1990){Karimabadi}, {Akimoto}, {Omidi} and
  {Menyuk}}]{Karimabadi90:waves}
\bibinfo{author}{{Karimabadi}, H.}, \bibinfo{author}{{Akimoto}, K.},
  \bibinfo{author}{{Omidi}, N.}, \bibinfo{author}{{Menyuk}, C.R.},
  \bibinfo{year}{1990}.
\newblock \bibinfo{title}{{Particle acceleration by a wave in a strong magnetic
  field - Regular and stochastic motion}}.
\newblock \bibinfo{journal}{Physics of Fluids B} \bibinfo{volume}{2},
  \bibinfo{pages}{606--628}.
\newblock \DOIprefix\doi{10.1063/1.859296}.
\bibitem[{{Karpman}(1974)}]{Karpman74:ssr}
\bibinfo{author}{{Karpman}, V.I.}, \bibinfo{year}{1974}.
\newblock \bibinfo{title}{{Nonlinear Effects in the ELF Waves Propagating along
  the Magnetic Field in the Magnetosphere}}.
\newblock \bibinfo{journal}{\ssr} \bibinfo{volume}{16},
  \bibinfo{pages}{361--388}.
\newblock \DOIprefix\doi{10.1007/BF00171564}.
\bibitem[{{Karpman} et~al.(1975){Karpman}, {Istomin} and {Shklyar}}]{Karpman75}
\bibinfo{author}{{Karpman}, V.I.}, \bibinfo{author}{{Istomin}, J.N.},
  \bibinfo{author}{{Shklyar}, D.R.}, \bibinfo{year}{1975}.
\newblock \bibinfo{title}{{Particle acceleration by a non-linear langmuir wave
  in an inhomogeneous plasma}}.
\newblock \bibinfo{journal}{Physics Letters A} \bibinfo{volume}{53},
  \bibinfo{pages}{101--102}.
\newblock \DOIprefix\doi{10.1016/0375-9601(75)90364-3}.
\bibitem[{{Kennel} and {Engelmann}(1966)}]{Kennel&Engelmann66}
\bibinfo{author}{{Kennel}, C.F.}, \bibinfo{author}{{Engelmann}, F.},
  \bibinfo{year}{1966}.
\newblock \bibinfo{title}{{Velocity Space Diffusion from Weak Plasma Turbulence
  in a Magnetic Field}}.
\newblock \bibinfo{journal}{Physics of Fluids} \bibinfo{volume}{9},
  \bibinfo{pages}{2377--2388}.
\newblock \DOIprefix\doi{10.1063/1.1761629}.
\bibitem[{{Le Contel} et~al.(2009){Le Contel}, {Roux}, {Jacquey}, {Robert},
  {Berthomier}, {Chust}, {Grison}, {Angelopoulos}, {Sibeck}, {Chaston},
  {Cully}, {Ergun}, {Glassmeier}, {Auster}, {McFadden}, {Carlson}, {Larson},
  {Bonnell}, {Mende}, {Russell}, {Donovan}, {Mann} and {Singer}}]{LeContel09}
\bibinfo{author}{{Le Contel}, O.}, \bibinfo{author}{{Roux}, A.},
  \bibinfo{author}{{Jacquey}, C.}, \bibinfo{author}{{Robert}, P.},
  \bibinfo{author}{{Berthomier}, M.}, \bibinfo{author}{{Chust}, T.},
  \bibinfo{author}{{Grison}, B.}, \bibinfo{author}{{Angelopoulos}, V.},
  \bibinfo{author}{{Sibeck}, D.}, \bibinfo{author}{{Chaston}, C.C.},
  \bibinfo{author}{{Cully}, C.M.}, \bibinfo{author}{{Ergun}, B.},
  \bibinfo{author}{{Glassmeier}, K.H.}, \bibinfo{author}{{Auster}, U.},
  \bibinfo{author}{{McFadden}, J.}, \bibinfo{author}{{Carlson}, C.},
  \bibinfo{author}{{Larson}, D.}, \bibinfo{author}{{Bonnell}, J.W.},
  \bibinfo{author}{{Mende}, S.}, \bibinfo{author}{{Russell}, C.T.},
  \bibinfo{author}{{Donovan}, E.}, \bibinfo{author}{{Mann}, I.},
  \bibinfo{author}{{Singer}, H.}, \bibinfo{year}{2009}.
\newblock \bibinfo{title}{{Quasi-parallel whistler mode waves observed by
  THEMIS during near-earth dipolarizations}}.
\newblock \bibinfo{journal}{Annales Geophysicae} \bibinfo{volume}{27},
  \bibinfo{pages}{2259--2275}.
\newblock \DOIprefix\doi{10.5194/angeo-27-2259-2009}.
\bibitem[{{Lerche}(1968)}]{Lerche68}
\bibinfo{author}{{Lerche}, I.}, \bibinfo{year}{1968}.
\newblock \bibinfo{title}{{Quasilinear Theory of Resonant Diffusion in a
  Magneto-Active, Relativistic Plasma}}.
\newblock \bibinfo{journal}{Physics of Fluids} \bibinfo{volume}{11}.
\bibitem[{{Li} et~al.(2011){Li}, {Bortnik}, {Thorne} and {Angelopoulos}}]{Li11}
\bibinfo{author}{{Li}, W.}, \bibinfo{author}{{Bortnik}, J.},
  \bibinfo{author}{{Thorne}, R.M.}, \bibinfo{author}{{Angelopoulos}, V.},
  \bibinfo{year}{2011}.
\newblock \bibinfo{title}{{Global distribution of wave amplitudes and wave
  normal angles of chorus waves using THEMIS wave observations}}.
\newblock \bibinfo{journal}{\jgr} \bibinfo{volume}{116},
  \bibinfo{pages}{12205}.
\newblock \DOIprefix\doi{10.1029/2011JA017035}.
\bibitem[{{Li} and {Hudson}(2019)}]{Li&Hudson19}
\bibinfo{author}{{Li}, W.}, \bibinfo{author}{{Hudson}, M.K.},
  \bibinfo{year}{2019}.
\newblock \bibinfo{title}{{Earth's Van Allen Radiation Belts: From Discovery to
  the Van Allen Probes Era}}.
\newblock \bibinfo{journal}{Journal of Geophysical Research (Space Physics)}
  \bibinfo{volume}{124}, \bibinfo{pages}{8319--8351}.
\newblock \DOIprefix\doi{10.1029/2018JA025940}.
\bibitem[{{Li} et~al.(2016a){Li}, {Mourenas}, {Artemyev}, {Bortnik}, {Thorne},
  {Kletzing}, {Kurth}, {Hospodarsky}, {Reeves}, {Funsten} and {Spence}}]{Li16}
\bibinfo{author}{{Li}, W.}, \bibinfo{author}{{Mourenas}, D.},
  \bibinfo{author}{{Artemyev}, A.V.}, \bibinfo{author}{{Bortnik}, J.},
  \bibinfo{author}{{Thorne}, R.M.}, \bibinfo{author}{{Kletzing}, C.A.},
  \bibinfo{author}{{Kurth}, W.S.}, \bibinfo{author}{{Hospodarsky}, G.B.},
  \bibinfo{author}{{Reeves}, G.D.}, \bibinfo{author}{{Funsten}, H.O.},
  \bibinfo{author}{{Spence}, H.E.}, \bibinfo{year}{2016}a.
\newblock \bibinfo{title}{{Unraveling the excitation mechanisms of highly
  oblique lower band chorus waves}}.
\newblock \bibinfo{journal}{\grl} \bibinfo{volume}{43},
  \bibinfo{pages}{8867--8875}.
\newblock \DOIprefix\doi{10.1002/2016GL070386}.
\bibitem[{{Li} et~al.(2016b){Li}, {Santolik}, {Bortnik}, {Thorne}, {Kletzing},
  {Kurth} and {Hospodarsky}}]{Li16:statistics}
\bibinfo{author}{{Li}, W.}, \bibinfo{author}{{Santolik}, O.},
  \bibinfo{author}{{Bortnik}, J.}, \bibinfo{author}{{Thorne}, R.M.},
  \bibinfo{author}{{Kletzing}, C.A.}, \bibinfo{author}{{Kurth}, W.S.},
  \bibinfo{author}{{Hospodarsky}, G.B.}, \bibinfo{year}{2016}b.
\newblock \bibinfo{title}{{New chorus wave properties near the equator from Van
  Allen Probes wave observations}}.
\newblock \bibinfo{journal}{\grl} \bibinfo{volume}{43},
  \bibinfo{pages}{4725--4735}.
\newblock \DOIprefix\doi{10.1002/2016GL068780}.
\bibitem[{{Li} et~al.(2012){Li}, {Thorne}, {Bortnik}, {Tao} and
  {Angelopoulos}}]{Li12}
\bibinfo{author}{{Li}, W.}, \bibinfo{author}{{Thorne}, R.M.},
  \bibinfo{author}{{Bortnik}, J.}, \bibinfo{author}{{Tao}, X.},
  \bibinfo{author}{{Angelopoulos}, V.}, \bibinfo{year}{2012}.
\newblock \bibinfo{title}{{Characteristics of hiss-like and discrete
  whistler-mode emissions}}.
\newblock \bibinfo{journal}{\grl} \bibinfo{volume}{39}, \bibinfo{pages}{18106}.
\newblock \DOIprefix\doi{10.1029/2012GL053206}.
\bibitem[{{Liu} et~al.(2019){Liu}, {Angelopoulos} and {Lu}}]{Liu19:foreshock}
\bibinfo{author}{{Liu}, T.Z.}, \bibinfo{author}{{Angelopoulos}, V.},
  \bibinfo{author}{{Lu}, S.}, \bibinfo{year}{2019}.
\newblock \bibinfo{title}{{Relativistic electrons generated at Earth's
  quasi-parallel bow shock}}.
\newblock \bibinfo{journal}{Science Advances} \bibinfo{volume}{5},
  \bibinfo{pages}{eaaw1368}.
\newblock \DOIprefix\doi{10.1126/sciadv.aaw1368}.
\bibitem[{{Lyons} and {Williams}(1984)}]{bookLyons&Williams}
\bibinfo{author}{{Lyons}, L.R.}, \bibinfo{author}{{Williams}, D.J.},
  \bibinfo{year}{1984}.
\newblock \bibinfo{title}{{Quantitative aspects of magnetospheric physics.}}
\bibitem[{{Menietti} et~al.(2012){Menietti}, {Shprits}, {Horne}, {Woodfield},
  {Hospodarsky} and {Gurnett}}]{Menietti12}
\bibinfo{author}{{Menietti}, J.D.}, \bibinfo{author}{{Shprits}, Y.Y.},
  \bibinfo{author}{{Horne}, R.B.}, \bibinfo{author}{{Woodfield}, E.E.},
  \bibinfo{author}{{Hospodarsky}, G.B.}, \bibinfo{author}{{Gurnett}, D.A.},
  \bibinfo{year}{2012}.
\newblock \bibinfo{title}{{Chorus, ECH, and Z mode emissions observed at
  Jupiter and Saturn and possible electron acceleration}}.
\newblock \bibinfo{journal}{\jgr} \bibinfo{volume}{117},
  \bibinfo{pages}{A12214}.
\newblock \DOIprefix\doi{10.1029/2012JA018187}.
\bibitem[{{Meredith} et~al.(2012){Meredith}, {Horne}, {Sicard-Piet}, {Boscher},
  {Yearby}, {Li} and {Thorne}}]{Meredith12}
\bibinfo{author}{{Meredith}, N.P.}, \bibinfo{author}{{Horne}, R.B.},
  \bibinfo{author}{{Sicard-Piet}, A.}, \bibinfo{author}{{Boscher}, D.},
  \bibinfo{author}{{Yearby}, K.H.}, \bibinfo{author}{{Li}, W.},
  \bibinfo{author}{{Thorne}, R.M.}, \bibinfo{year}{2012}.
\newblock \bibinfo{title}{{Global model of lower band and upper band chorus
  from multiple satellite observations}}.
\newblock \bibinfo{journal}{\jgr} \bibinfo{volume}{117},
  \bibinfo{pages}{10225}.
\newblock \DOIprefix\doi{10.1029/2012JA017978}.
\bibitem[{{Mourenas} et~al.(2018){Mourenas}, {Zhang}, {Artemyev},
  {Angelopoulos}, {Thorne}, {Bortnik}, {Neishtadt} and
  {Vasiliev}}]{Mourenas18:jgr}
\bibinfo{author}{{Mourenas}, D.}, \bibinfo{author}{{Zhang}, X.J.},
  \bibinfo{author}{{Artemyev}, A.V.}, \bibinfo{author}{{Angelopoulos}, V.},
  \bibinfo{author}{{Thorne}, R.M.}, \bibinfo{author}{{Bortnik}, J.},
  \bibinfo{author}{{Neishtadt}, A.I.}, \bibinfo{author}{{Vasiliev}, A.A.},
  \bibinfo{year}{2018}.
\newblock \bibinfo{title}{{Electron Nonlinear Resonant Interaction With Short
  and Intense Parallel Chorus Wave Packets}}.
\newblock \bibinfo{journal}{\jgr} \bibinfo{volume}{123},
  \bibinfo{pages}{4979--4999}.
\newblock \DOIprefix\doi{10.1029/2018JA025417}.
\bibitem[{{Oka} et~al.(2019){Oka}, {Otsuka}, {Matsukiyo}, {Wilson}, {Argall},
  {Amano}, {Phan}, {Hoshino}, {Le Contel}, {Gershman}, {Burch}, {Torbert},
  {Dorelli}, {Giles}, {Ergun}, {Russell} and {Lindqvist}}]{Oka19}
\bibinfo{author}{{Oka}, M.}, \bibinfo{author}{{Otsuka}, F.},
  \bibinfo{author}{{Matsukiyo}, S.}, \bibinfo{author}{{Wilson}, L.~B., I.},
  \bibinfo{author}{{Argall}, M.R.}, \bibinfo{author}{{Amano}, T.},
  \bibinfo{author}{{Phan}, T.D.}, \bibinfo{author}{{Hoshino}, M.},
  \bibinfo{author}{{Le Contel}, O.}, \bibinfo{author}{{Gershman}, D.J.},
  \bibinfo{author}{{Burch}, J.L.}, \bibinfo{author}{{Torbert}, R.B.},
  \bibinfo{author}{{Dorelli}, J.C.}, \bibinfo{author}{{Giles}, B.L.},
  \bibinfo{author}{{Ergun}, R.E.}, \bibinfo{author}{{Russell}, C.T.},
  \bibinfo{author}{{Lindqvist}, P.A.}, \bibinfo{year}{2019}.
\newblock \bibinfo{title}{{Electron Scattering by Low-frequency Whistler Waves
  at Earth{\textquoteright}s Bow Shock}}.
\newblock \bibinfo{journal}{\apj} \bibinfo{volume}{886}, \bibinfo{pages}{53}.
\newblock \DOIprefix\doi{10.3847/1538-4357/ab4a81}.
\bibitem[{{Oka} et~al.(2017){Oka}, {Wilson}, {Phan}, {Hull}, {Amano},
  {Hoshino}, {Argall}, {Le Contel}, {Agapitov}, {Gershman}, {Khotyaintsev},
  {Burch}, {Torbert}, {Pollock}, {Dorelli}, {Giles}, {Moore}, {Saito},
  {Avanov}, {Paterson}, {Ergun}, {Strangeway}, {Russell} and
  {Lindqvist}}]{Oka17}
\bibinfo{author}{{Oka}, M.}, \bibinfo{author}{{Wilson}, III, L.B.},
  \bibinfo{author}{{Phan}, T.D.}, \bibinfo{author}{{Hull}, A.J.},
  \bibinfo{author}{{Amano}, T.}, \bibinfo{author}{{Hoshino}, M.},
  \bibinfo{author}{{Argall}, M.R.}, \bibinfo{author}{{Le Contel}, O.},
  \bibinfo{author}{{Agapitov}, O.}, \bibinfo{author}{{Gershman}, D.J.},
  \bibinfo{author}{{Khotyaintsev}, Y.V.}, \bibinfo{author}{{Burch}, J.L.},
  \bibinfo{author}{{Torbert}, R.B.}, \bibinfo{author}{{Pollock}, C.},
  \bibinfo{author}{{Dorelli}, J.C.}, \bibinfo{author}{{Giles}, B.L.},
  \bibinfo{author}{{Moore}, T.E.}, \bibinfo{author}{{Saito}, Y.},
  \bibinfo{author}{{Avanov}, L.A.}, \bibinfo{author}{{Paterson}, W.},
  \bibinfo{author}{{Ergun}, R.E.}, \bibinfo{author}{{Strangeway}, R.J.},
  \bibinfo{author}{{Russell}, C.T.}, \bibinfo{author}{{Lindqvist}, P.A.},
  \bibinfo{year}{2017}.
\newblock \bibinfo{title}{{Electron Scattering by High-frequency Whistler Waves
  at Earth's Bow Shock}}.
\newblock \bibinfo{journal}{\apjl} \bibinfo{volume}{842}, \bibinfo{pages}{L11}.
\newblock \DOIprefix\doi{10.3847/2041-8213/aa7759}.
\bibitem[{{Omura} et~al.(2015){Omura}, {Miyashita}, {Yoshikawa}, {Summers},
  {Hikishima}, {Ebihara} and {Kubota}}]{Omura15}
\bibinfo{author}{{Omura}, Y.}, \bibinfo{author}{{Miyashita}, Y.},
  \bibinfo{author}{{Yoshikawa}, M.}, \bibinfo{author}{{Summers}, D.},
  \bibinfo{author}{{Hikishima}, M.}, \bibinfo{author}{{Ebihara}, Y.},
  \bibinfo{author}{{Kubota}, Y.}, \bibinfo{year}{2015}.
\newblock \bibinfo{title}{{Formation process of relativistic electron flux
  through interaction with chorus emissions in the Earth's inner
  magnetosphere}}.
\newblock \bibinfo{journal}{\jgr} \bibinfo{volume}{120},
  \bibinfo{pages}{9545--9562}.
\newblock \DOIprefix\doi{10.1002/2015JA021563}.
\bibitem[{{Schulz} and {Lanzerotti}(1974)}]{bookSchulz&anzerotti74}
\bibinfo{author}{{Schulz}, M.}, \bibinfo{author}{{Lanzerotti}, L.J.},
  \bibinfo{year}{1974}.
\newblock \bibinfo{title}{{Particle diffusion in the radiation belts}}.
\newblock \bibinfo{publisher}{Springer, New York}.
\bibitem[{{Schure} et~al.(2012){Schure}, {Bell}, {O'C Drury} and
  {Bykov}}]{Schure12:diffusive_shock_acceleration}
\bibinfo{author}{{Schure}, K.M.}, \bibinfo{author}{{Bell}, A.R.},
  \bibinfo{author}{{O'C Drury}, L.}, \bibinfo{author}{{Bykov}, A.M.},
  \bibinfo{year}{2012}.
\newblock \bibinfo{title}{{Diffusive Shock Acceleration and Magnetic Field
  Amplification}}.
\newblock \bibinfo{journal}{\ssr} \bibinfo{volume}{173},
  \bibinfo{pages}{491--519}.
\newblock \DOIprefix\doi{10.1007/s11214-012-9871-7},
  \href{http://arxiv.org/abs/1203.1637}{{\tt arXiv:1203.1637}}.
\bibitem[{{Schwartz} et~al.(2011){Schwartz}, {Henley}, {Mitchell} and
  {Krasnoselskikh}}]{Schwartz11:VladimirShock}
\bibinfo{author}{{Schwartz}, S.J.}, \bibinfo{author}{{Henley}, E.},
  \bibinfo{author}{{Mitchell}, J.}, \bibinfo{author}{{Krasnoselskikh}, V.},
  \bibinfo{year}{2011}.
\newblock \bibinfo{title}{{Electron Temperature Gradient Scale at Collisionless
  Shocks}}.
\newblock \bibinfo{journal}{Physical Review Letters} \bibinfo{volume}{107},
  \bibinfo{pages}{215002}.
\newblock \DOIprefix\doi{10.1103/PhysRevLett.107.215002}.
\bibitem[{{Shaaban} et~al.(2019){Shaaban}, {Lazar}, {Yoon}, {Poedts} and
  {L{\'o}pez}}]{Shaaban19}
\bibinfo{author}{{Shaaban}, S.M.}, \bibinfo{author}{{Lazar}, M.},
  \bibinfo{author}{{Yoon}, P.H.}, \bibinfo{author}{{Poedts}, S.},
  \bibinfo{author}{{L{\'o}pez}, R.A.}, \bibinfo{year}{2019}.
\newblock \bibinfo{title}{{Quasi-linear approach of the whistler heat-flux
  instability in the solar wind}}.
\newblock \bibinfo{journal}{\mnras} \bibinfo{volume}{486},
  \bibinfo{pages}{4498--4507}.
\newblock \DOIprefix\doi{10.1093/mnras/stz830},
  \href{http://arxiv.org/abs/1903.08005}{{\tt arXiv:1903.08005}}.
\bibitem[{{Shapiro} and {Sagdeev}(1997)}]{Shapiro&Sagdeev97}
\bibinfo{author}{{Shapiro}, V.D.}, \bibinfo{author}{{Sagdeev}, R.Z.},
  \bibinfo{year}{1997}.
\newblock \bibinfo{title}{{Nonlinear wave-particle interaction and conditions
  for the applicability of quasilinear theory}}.
\newblock \bibinfo{journal}{Physics Reports} \bibinfo{volume}{283},
  \bibinfo{pages}{49--71}.
\newblock \DOIprefix\doi{10.1016/S0370-1573(96)00053-1}.
\bibitem[{{Sheeley} et~al.(2001){Sheeley}, {Moldwin}, {Rassoul} and
  {Anderson}}]{Sheeley01}
\bibinfo{author}{{Sheeley}, B.W.}, \bibinfo{author}{{Moldwin}, M.B.},
  \bibinfo{author}{{Rassoul}, H.K.}, \bibinfo{author}{{Anderson}, R.R.},
  \bibinfo{year}{2001}.
\newblock \bibinfo{title}{{An empirical plasmasphere and trough density model:
  CRRES observations}}.
\newblock \bibinfo{journal}{\jgr} \bibinfo{volume}{106},
  \bibinfo{pages}{25631--25642}.
\newblock \DOIprefix\doi{10.1029/2000JA000286}.
\bibitem[{{Shen} et~al.(2020){Shen}, {Artemyev}, {Zhang}, {Vasko}, {Runov},
  {Angelopoulos} and {Knudsen}}]{Shen20:tds}
\bibinfo{author}{{Shen}, Y.}, \bibinfo{author}{{Artemyev}, A.},
  \bibinfo{author}{{Zhang}, X.J.}, \bibinfo{author}{{Vasko}, I.Y.},
  \bibinfo{author}{{Runov}, A.}, \bibinfo{author}{{Angelopoulos}, V.},
  \bibinfo{author}{{Knudsen}, D.}, \bibinfo{year}{2020}.
\newblock \bibinfo{title}{{Potential Evidence of Low-Energy Electron Scattering
  and Ionospheric Precipitation by Time Domain Structures}}.
\newblock \bibinfo{journal}{\grl} \bibinfo{volume}{47},
  \bibinfo{pages}{e89138}.
\newblock \DOIprefix\doi{10.1029/2020GL089138}.
\bibitem[{{Shklyar}(1981)}]{Shklyar81}
\bibinfo{author}{{Shklyar}, D.R.}, \bibinfo{year}{1981}.
\newblock \bibinfo{title}{{Stochastic motion of relativistic particles in the
  field of a monochromatic wave}}.
\newblock \bibinfo{journal}{Sov. Phys. JETP} \bibinfo{volume}{53},
  \bibinfo{pages}{1197--1192}.
\bibitem[{{Shklyar}(2011)}]{Shklyar11:angeo}
\bibinfo{author}{{Shklyar}, D.R.}, \bibinfo{year}{2011}.
\newblock \bibinfo{title}{{On the nature of particle energization via resonant
  wave-particle interaction in the inhomogeneous magnetospheric plasma}}.
\newblock \bibinfo{journal}{Annales Geophysicae} \bibinfo{volume}{29},
  \bibinfo{pages}{1179--1188}.
\newblock \DOIprefix\doi{10.5194/angeo-29-1179-2011}.
\bibitem[{{Shklyar} et~al.(2020){Shklyar}, {Manninen}, {Titova}, {Santolík},
  {Kolmašová} and {Turunen}}]{Shklyar20}
\bibinfo{author}{{Shklyar}, D.R.}, \bibinfo{author}{{Manninen}, J.},
  \bibinfo{author}{{Titova}, E.}, \bibinfo{author}{{Santolík}, O.},
  \bibinfo{author}{{Kolmašová}, I.}, \bibinfo{author}{{Turunen}, T.},
  \bibinfo{year}{2020}.
\newblock \bibinfo{title}{Ground and space signatures of vlf noise suppression
  by whistlers}.
\newblock \bibinfo{journal}{Journal of Geophysical Research: Space Physics} ,
  \bibinfo{pages}{e2019JA027430}\DOIprefix\doi{10.1029/2019JA027430}.
\bibitem[{{Shklyar} and {Matsumoto}(2009)}]{Shklyar09:review}
\bibinfo{author}{{Shklyar}, D.R.}, \bibinfo{author}{{Matsumoto}, H.},
  \bibinfo{year}{2009}.
\newblock \bibinfo{title}{{Oblique Whistler-Mode Waves in the Inhomogeneous
  Magnetospheric Plasma: Resonant Interactions with Energetic Charged
  Particles}}.
\newblock \bibinfo{journal}{Surveys in Geophysics} \bibinfo{volume}{30},
  \bibinfo{pages}{55--104}.
\newblock \DOIprefix\doi{10.1007/s10712-009-9061-7}.
\bibitem[{{Shklyar} and {Zimbardo}(2014)}]{Shklyar&Zimbardo14}
\bibinfo{author}{{Shklyar}, D.R.}, \bibinfo{author}{{Zimbardo}, G.},
  \bibinfo{year}{2014}.
\newblock \bibinfo{title}{{Particle dynamics in the field of two waves in a
  magnetoplasma}}.
\newblock \bibinfo{journal}{Plasma Physics and Controlled Fusion}
  \bibinfo{volume}{56}, \bibinfo{pages}{095002}.
\newblock \DOIprefix\doi{10.1088/0741-3335/56/9/095002}.
\bibitem[{{Shprits} et~al.(2008){Shprits}, {Subbotin}, {Meredith} and
  {Elkington}}]{Shprits08:JASTP_local}
\bibinfo{author}{{Shprits}, Y.Y.}, \bibinfo{author}{{Subbotin}, D.A.},
  \bibinfo{author}{{Meredith}, N.P.}, \bibinfo{author}{{Elkington}, S.R.},
  \bibinfo{year}{2008}.
\newblock \bibinfo{title}{{Review of modeling of losses and sources of
  relativistic electrons in the outer radiation belt II: Local acceleration and
  loss}}.
\newblock \bibinfo{journal}{Journal of Atmospheric and Solar-Terrestrial
  Physics} \bibinfo{volume}{70}, \bibinfo{pages}{1694--1713}.
\newblock \DOIprefix\doi{10.1016/j.jastp.2008.06.014}.
\bibitem[{{Solovev} and {Shkliar}(1986)}]{Solovev&Shkliar86}
\bibinfo{author}{{Solovev}, V.V.}, \bibinfo{author}{{Shkliar}, D.R.},
  \bibinfo{year}{1986}.
\newblock \bibinfo{title}{{Particle heating by a low-amplitude wave in an
  inhomogeneous magnetoplasma}}.
\newblock \bibinfo{journal}{Sov. Phys. JETP} \bibinfo{volume}{63},
  \bibinfo{pages}{272--277}.
\bibitem[{{Sorathia} et~al.(2017){Sorathia}, {Merkin}, {Ukhorskiy}, {Mauk} and
  {Sibeck}}]{Sorathia17}
\bibinfo{author}{{Sorathia}, K.A.}, \bibinfo{author}{{Merkin}, V.G.},
  \bibinfo{author}{{Ukhorskiy}, A.Y.}, \bibinfo{author}{{Mauk}, B.H.},
  \bibinfo{author}{{Sibeck}, D.G.}, \bibinfo{year}{2017}.
\newblock \bibinfo{title}{{Energetic particle loss through the magnetopause: A
  combined global MHD and test-particle study}}.
\newblock \bibinfo{journal}{Journal of Geophysical Research (Space Physics)}
  \bibinfo{volume}{122}, \bibinfo{pages}{9329--9343}.
\newblock \DOIprefix\doi{10.1002/2017JA024268}.
\bibitem[{{Sorathia} et~al.(2018){Sorathia}, {Ukhorskiy}, {Merkin}, {Fennell}
  and {Claudepierre}}]{Sorathia18}
\bibinfo{author}{{Sorathia}, K.A.}, \bibinfo{author}{{Ukhorskiy}, A.Y.},
  \bibinfo{author}{{Merkin}, V.G.}, \bibinfo{author}{{Fennell}, J.F.},
  \bibinfo{author}{{Claudepierre}, S.G.}, \bibinfo{year}{2018}.
\newblock \bibinfo{title}{{Modeling the Depletion and Recovery of the Outer
  Radiation Belt During a Geomagnetic Storm: Combined MHD and Test Particle
  Simulations}}.
\newblock \bibinfo{journal}{Journal of Geophysical Research (Space Physics)}
  \bibinfo{volume}{123}, \bibinfo{pages}{5590--5609}.
\newblock \DOIprefix\doi{10.1029/2018JA025506}.
\bibitem[{{Stix}(1962)}]{bookStix62}
\bibinfo{author}{{Stix}, T.H.}, \bibinfo{year}{1962}.
\newblock \bibinfo{title}{{The Theory of Plasma Waves}}.
\bibitem[{{Tao} et~al.(2008){Tao}, {Chan}, {Albert} and
  {Miller}}]{Tao08:stochastic}
\bibinfo{author}{{Tao}, X.}, \bibinfo{author}{{Chan}, A.A.},
  \bibinfo{author}{{Albert}, J.M.}, \bibinfo{author}{{Miller}, J.A.},
  \bibinfo{year}{2008}.
\newblock \bibinfo{title}{{Stochastic modeling of multidimensional diffusion in
  the radiation belts}}.
\newblock \bibinfo{journal}{Journal of Geophysical Research (Space Physics)}
  \bibinfo{volume}{113}, \bibinfo{pages}{A07212}.
\newblock \DOIprefix\doi{10.1029/2007JA012985}.
\bibitem[{{Thorne}(2010)}]{Thorne10:GRL}
\bibinfo{author}{{Thorne}, R.M.}, \bibinfo{year}{2010}.
\newblock \bibinfo{title}{{Radiation belt dynamics: The importance of
  wave-particle interactions}}.
\newblock \bibinfo{journal}{\grl} \bibinfo{volume}{372},
  \bibinfo{pages}{22107}.
\newblock \DOIprefix\doi{10.1029/2010GL044990}.
\bibitem[{{Thorne} et~al.(2013){Thorne}, {Li}, {Ni}, {Ma}, {Bortnik}, {Chen},
  {Baker}, {Spence}, {Reeves}, {Henderson}, {Kletzing}, {Kurth}, {Hospodarsky},
  {Blake}, {Fennell}, {Claudepierre} and {Kanekal}}]{Thorne13:nature}
\bibinfo{author}{{Thorne}, R.M.}, \bibinfo{author}{{Li}, W.},
  \bibinfo{author}{{Ni}, B.}, \bibinfo{author}{{Ma}, Q.},
  \bibinfo{author}{{Bortnik}, J.}, \bibinfo{author}{{Chen}, L.},
  \bibinfo{author}{{Baker}, D.N.}, \bibinfo{author}{{Spence}, H.E.},
  \bibinfo{author}{{Reeves}, G.D.}, \bibinfo{author}{{Henderson}, M.G.},
  \bibinfo{author}{{Kletzing}, C.A.}, \bibinfo{author}{{Kurth}, W.S.},
  \bibinfo{author}{{Hospodarsky}, G.B.}, \bibinfo{author}{{Blake}, J.B.},
  \bibinfo{author}{{Fennell}, J.F.}, \bibinfo{author}{{Claudepierre}, S.G.},
  \bibinfo{author}{{Kanekal}, S.G.}, \bibinfo{year}{2013}.
\newblock \bibinfo{title}{{Rapid local acceleration of relativistic
  radiation-belt electrons by magnetospheric chorus}}.
\newblock \bibinfo{journal}{\nat} \bibinfo{volume}{504},
  \bibinfo{pages}{411--414}.
\newblock \DOIprefix\doi{10.1038/nature12889}.
\bibitem[{{Tobita} and {Omura}(2018)}]{Tobita&Omura18}
\bibinfo{author}{{Tobita}, M.}, \bibinfo{author}{{Omura}, Y.},
  \bibinfo{year}{2018}.
\newblock \bibinfo{title}{{Nonlinear dynamics of resonant electrons interacting
  with coherent Langmuir waves}}.
\newblock \bibinfo{journal}{Physics of Plasmas} \bibinfo{volume}{25},
  \bibinfo{pages}{032105}.
\newblock \DOIprefix\doi{10.1063/1.5018084}.
\bibitem[{{Tong} et~al.(2019){Tong}, {Vasko}, {Artemyev}, {Bale} and
  {Mozer}}]{Tong19:ApJ}
\bibinfo{author}{{Tong}, Y.}, \bibinfo{author}{{Vasko}, I.Y.},
  \bibinfo{author}{{Artemyev}, A.V.}, \bibinfo{author}{{Bale}, S.D.},
  \bibinfo{author}{{Mozer}, F.S.}, \bibinfo{year}{2019}.
\newblock \bibinfo{title}{{Statistical Study of Whistler Waves in the Solar
  Wind at 1 au}}.
\newblock \bibinfo{journal}{\apj} \bibinfo{volume}{878}, \bibinfo{pages}{41}.
\newblock \DOIprefix\doi{10.3847/1538-4357/ab1f05},
  \href{http://arxiv.org/abs/1905.08958}{{\tt arXiv:1905.08958}}.
\bibitem[{{Trakhtengerts} and {Rycroft}(2008)}]{bookTrakhtengerts&Rycroft08}
\bibinfo{author}{{Trakhtengerts}, V.Y.}, \bibinfo{author}{{Rycroft}, M.J.},
  \bibinfo{year}{2008}.
\newblock \bibinfo{title}{{Whistler and Alfv{\'e}n Mode Cyclotron Masers in
  Space}}.
\newblock \bibinfo{publisher}{Cambridge University Press}.
\bibitem[{{Tsurutani} et~al.(2020){Tsurutani}, {Chen}, {Gao}, {Lu}, {Pickett},
  {Lakhina}, {Sen}, {Hajra}, {Park} and {Falkowski}}]{Tsurutani20:jgr}
\bibinfo{author}{{Tsurutani}, B.T.}, \bibinfo{author}{{Chen}, R.},
  \bibinfo{author}{{Gao}, X.}, \bibinfo{author}{{Lu}, Q.},
  \bibinfo{author}{{Pickett}, J.S.}, \bibinfo{author}{{Lakhina}, G.S.},
  \bibinfo{author}{{Sen}, A.}, \bibinfo{author}{{Hajra}, R.},
  \bibinfo{author}{{Park}, S.A.}, \bibinfo{author}{{Falkowski}, B.J.},
  \bibinfo{year}{2020}.
\newblock \bibinfo{title}{Lower-band “monochromatic” chorus riser
  subelement/wave packet observations}.
\newblock \bibinfo{journal}{Journal of Geophysical Research: Space Physics} ,
  \bibinfo{pages}{e2020JA028090}
  \DOIprefix\doi{10.1029/2020JA028090}.
\bibitem[{Turner et~al.(2016)Turner, Fennell, Blake, Clemmons, Mauk, Cohen,
  Jaynes, Craft, Wilder, Baker, Reeves, Gershman, Avanov, Dorelli, Giles,
  Pollock, Schmid, Nakamura, Strangeway, Russell, Artemyev, Runov,
  Angelopoulos, Spence, Torbert and Burch}]{Turner16}
\bibinfo{author}{Turner, D.L.}, \bibinfo{author}{Fennell, J.F.},
  \bibinfo{author}{Blake, J.B.}, \bibinfo{author}{Clemmons, J.H.},
  \bibinfo{author}{Mauk, B.H.}, \bibinfo{author}{Cohen, I.J.},
  \bibinfo{author}{Jaynes, A.N.}, \bibinfo{author}{Craft, J.V.},
  \bibinfo{author}{Wilder, F.D.}, \bibinfo{author}{Baker, D.N.},
  \bibinfo{author}{Reeves, G.D.}, \bibinfo{author}{Gershman, D.J.},
  \bibinfo{author}{Avanov, L.A.}, \bibinfo{author}{Dorelli, J.C.},
  \bibinfo{author}{Giles, B.L.}, \bibinfo{author}{Pollock, C.J.},
  \bibinfo{author}{Schmid, D.}, \bibinfo{author}{Nakamura, R.},
  \bibinfo{author}{Strangeway, R.J.}, \bibinfo{author}{Russell, C.T.},
  \bibinfo{author}{Artemyev, A.V.}, \bibinfo{author}{Runov, A.},
  \bibinfo{author}{Angelopoulos, V.}, \bibinfo{author}{Spence, H.E.},
  \bibinfo{author}{Torbert, R.B.}, \bibinfo{author}{Burch, J.L.},
  \bibinfo{year}{2016}.
\newblock \bibinfo{title}{Energy limits of electron acceleration in the plasma
  sheet during substorms: A case study with the magnetospheric multiscale (mms)
  mission}.
\newblock \bibinfo{journal}{\grl} \bibinfo{volume}{43},
  \bibinfo{pages}{7785--7794}.
\newblock \URLprefix \url{http://dx.doi.org/10.1002/2016GL069691},
  \DOIprefix\doi{10.1002/2016GL069691}.
\bibitem[{{Vainchtein} et~al.(2017){Vainchtein}, {Fridman} and
  {Artemyev}}]{Vainchtein17}
\bibinfo{author}{{Vainchtein}, D.}, \bibinfo{author}{{Fridman}, G.},
  \bibinfo{author}{{Artemyev}, A.}, \bibinfo{year}{2017}.
\newblock \bibinfo{title}{{Generation of discrete structures in phase-space via
  charged particle trapping by an electrostatic wave}}.
\newblock \bibinfo{journal}{Communications in Nonlinear Science and Numerical
  Simulations} \bibinfo{volume}{51}, \bibinfo{pages}{133--140}.
\newblock \DOIprefix\doi{10.1016/j.cnsns.2017.04.003}.
\bibitem[{{Vainchtein} et~al.(2018){Vainchtein}, {Zhang}, {Artemyev},
  {Mourenas}, {Angelopoulos} and {Thorne}}]{Vainchtein18:jgr}
\bibinfo{author}{{Vainchtein}, D.}, \bibinfo{author}{{Zhang}, X.J.},
  \bibinfo{author}{{Artemyev}, A.V.}, \bibinfo{author}{{Mourenas}, D.},
  \bibinfo{author}{{Angelopoulos}, V.}, \bibinfo{author}{{Thorne}, R.M.},
  \bibinfo{year}{2018}.
\newblock \bibinfo{title}{{Evolution of Electron Distribution Driven by
  Nonlinear Resonances With Intense Field-Aligned Chorus Waves}}.
\newblock \bibinfo{journal}{Journal of Geophysical Research (Space Physics)}
  \bibinfo{volume}{123}, \bibinfo{pages}{8149--8169}.
\newblock \DOIprefix\doi{10.1029/2018JA025654},
  \href{http://arxiv.org/abs/1806.00066}{{\tt arXiv:1806.00066}}.
\bibitem[{{Vasko} et~al.(2017){Vasko}, {Agapitov}, {Mozer}, {Artemyev},
  {Krasnoselskikh} and {Bonnell}}]{Vasko17:diffusion}
\bibinfo{author}{{Vasko}, I.Y.}, \bibinfo{author}{{Agapitov}, O.V.},
  \bibinfo{author}{{Mozer}, F.S.}, \bibinfo{author}{{Artemyev}, A.V.},
  \bibinfo{author}{{Krasnoselskikh}, V.V.}, \bibinfo{author}{{Bonnell}, J.W.},
  \bibinfo{year}{2017}.
\newblock \bibinfo{title}{{Diffusive scattering of electrons by electron holes
  around injection fronts}}.
\newblock \bibinfo{journal}{\jgr} \bibinfo{volume}{122},
  \bibinfo{pages}{3163--3182}.
\newblock \DOIprefix\doi{10.1002/2016JA023337}.
\bibitem[{{Vasko} et~al.(2018){Vasko}, {Krasnoselskikh}, {Mozer} and
  {Artemyev}}]{Vasko18:pop}
\bibinfo{author}{{Vasko}, I.Y.}, \bibinfo{author}{{Krasnoselskikh}, V.V.},
  \bibinfo{author}{{Mozer}, F.S.}, \bibinfo{author}{{Artemyev}, A.V.},
  \bibinfo{year}{2018}.
\newblock \bibinfo{title}{{Scattering by the broadband electrostatic turbulence
  in the space plasma}}.
\newblock \bibinfo{journal}{Physics of Plasmas} \bibinfo{volume}{25},
  \bibinfo{pages}{072903}.
\newblock \DOIprefix\doi{10.1063/1.5039687}.
\bibitem[{{Vedenov} et~al.(1962){Vedenov}, {Velikhov} and
  {Sagdeev}}]{Vedenov62}
\bibinfo{author}{{Vedenov}, A.A.}, \bibinfo{author}{{Velikhov}, E.},
  \bibinfo{author}{{Sagdeev}, R.}, \bibinfo{year}{1962}.
\newblock \bibinfo{title}{{Quasilinear theory of plasma oscillations}}.
\newblock \bibinfo{journal}{Nuclear Fusion Suppl.} \bibinfo{volume}{2},
  \bibinfo{pages}{465--475}.
\bibitem[{{Vlahos} and {Isliker}(2019)}]{Vlahos&Isliker19}
\bibinfo{author}{{Vlahos}, L.}, \bibinfo{author}{{Isliker}, H.},
  \bibinfo{year}{2019}.
\newblock \bibinfo{title}{{Particle acceleration and heating in a turbulent
  solar corona}}.
\newblock \bibinfo{journal}{Plasma Physics and Controlled Fusion}
  \bibinfo{volume}{61}, \bibinfo{pages}{014020}.
\newblock \DOIprefix\doi{10.1088/1361-6587/aadbe7},
  \href{http://arxiv.org/abs/1808.07136}{{\tt arXiv:1808.07136}}.
\bibitem[{{Wilson} et~al.(2011){Wilson}, {Cattell}, {Kellogg}, {Wygant},
  {Goetz}, {Breneman} and {Kersten}}]{Cattell11:Wilson}
\bibinfo{author}{{Wilson}, III, L.B.}, \bibinfo{author}{{Cattell}, C.A.},
  \bibinfo{author}{{Kellogg}, P.J.}, \bibinfo{author}{{Wygant}, J.R.},
  \bibinfo{author}{{Goetz}, K.}, \bibinfo{author}{{Breneman}, A.},
  \bibinfo{author}{{Kersten}, K.}, \bibinfo{year}{2011}.
\newblock \bibinfo{title}{{The properties of large amplitude whistler mode
  waves in the magnetosphere: Propagation and relationship with geomagnetic
  activity}}.
\newblock \bibinfo{journal}{\grl} \bibinfo{volume}{38}, \bibinfo{pages}{17107}.
\newblock \DOIprefix\doi{10.1029/2011GL048671}.
\bibitem[{{Zhang} et~al.(2018a){Zhang}, {Angelopoulos}, {Artemyev} and
  {Liu}}]{Zhang18:whistlers&injections}
\bibinfo{author}{{Zhang}, X.}, \bibinfo{author}{{Angelopoulos}, V.},
  \bibinfo{author}{{Artemyev}, A.V.}, \bibinfo{author}{{Liu}, J.},
  \bibinfo{year}{2018}a.
\newblock \bibinfo{title}{{Whistler and Electron Firehose Instability Control
  of Electron Distributions in and Around Dipolarizing Flux Bundles}}.
\newblock \bibinfo{journal}{\grl} \bibinfo{volume}{45},
  \bibinfo{pages}{9380--9389}.
\newblock \DOIprefix\doi{10.1029/2018GL079613}.
\bibitem[{{Zhang} et~al.(2019a){Zhang}, {Angelopoulos}, {Artemyev} and
  {Liu}}]{Zhang19:grl:whistlers}
\bibinfo{author}{{Zhang}, X.}, \bibinfo{author}{{Angelopoulos}, V.},
  \bibinfo{author}{{Artemyev}, A.V.}, \bibinfo{author}{{Liu}, J.},
  \bibinfo{year}{2019}a.
\newblock \bibinfo{title}{{Energy Transport by Whistler Waves Around
  Dipolarizing Flux Bundles}}.
\newblock \bibinfo{journal}{\grl} \bibinfo{volume}{46},
  \bibinfo{pages}{11,718--11,727}.
\newblock \DOIprefix\doi{10.1029/2019GL084226}.
\bibitem[{{Zhang} et~al.(2020){Zhang}, {Agapitov}, {Artemyev}, {Mourenas},
  {Angelopoulos}, {Kurth}, {Bonnell} and {Hospodarsky}}]{Zhang20:grl:phase}
\bibinfo{author}{{Zhang}, X.J.}, \bibinfo{author}{{Agapitov}, O.},
  \bibinfo{author}{{Artemyev}, A.V.}, \bibinfo{author}{{Mourenas}, D.},
  \bibinfo{author}{{Angelopoulos}, V.}, \bibinfo{author}{{Kurth}, W.S.},
  \bibinfo{author}{{Bonnell}, J.W.}, \bibinfo{author}{{Hospodarsky}, G.B.},
  \bibinfo{year}{2020}.
\newblock \bibinfo{title}{{Phase Decoherence Within Intense Chorus Wave Packets
  Constrains the Efficiency of Nonlinear Resonant Electron Acceleration}}.
\newblock \bibinfo{journal}{\grl} \bibinfo{volume}{47},
  \bibinfo{pages}{e89807}.
\newblock \DOIprefix\doi{10.1029/2020GL089807}.
\bibitem[{{Zhang} et~al.(2019b){Zhang}, {Chen}, {Artemyev}, {Angelopoulos} and
  {Liu}}]{Zhang19:jgr:modulation}
\bibinfo{author}{{Zhang}, X.J.}, \bibinfo{author}{{Chen}, L.},
  \bibinfo{author}{{Artemyev}, A.V.}, \bibinfo{author}{{Angelopoulos}, V.},
  \bibinfo{author}{{Liu}, X.}, \bibinfo{year}{2019}b.
\newblock \bibinfo{title}{{Periodic Excitation of Chorus and ECH Waves
  Modulated by Ultralow Frequency Compressions}}.
\newblock \bibinfo{journal}{Journal of Geophysical Research (Space Physics)}
  \bibinfo{volume}{124}, \bibinfo{pages}{8535--8550}.
\newblock \DOIprefix\doi{10.1029/2019JA027201}.
\bibitem[{{Zhang} et~al.(2018b){Zhang}, {Thorne}, {Artemyev}, {Mourenas},
  {Angelopoulos}, {Bortnik}, {Kletzing}, {Kurth} and
  {Hospodarsky}}]{Zhang18:jgr:intensewaves}
\bibinfo{author}{{Zhang}, X.J.}, \bibinfo{author}{{Thorne}, R.},
  \bibinfo{author}{{Artemyev}, A.}, \bibinfo{author}{{Mourenas}, D.},
  \bibinfo{author}{{Angelopoulos}, V.}, \bibinfo{author}{{Bortnik}, J.},
  \bibinfo{author}{{Kletzing}, C.A.}, \bibinfo{author}{{Kurth}, W.S.},
  \bibinfo{author}{{Hospodarsky}, G.B.}, \bibinfo{year}{2018}b.
\newblock \bibinfo{title}{{Properties of Intense Field-Aligned Lower-Band
  Chorus Waves: Implications for Nonlinear Wave-Particle Interactions}}.
\newblock \bibinfo{journal}{Journal of Geophysical Research (Space Physics)}
  \bibinfo{volume}{123}, \bibinfo{pages}{5379--5393}.
\newblock \DOIprefix\doi{10.1029/2018JA025390}.
\bibitem[{{Zhang} et~al.(1999){Zhang}, {Matsumoto}, {Kojima} and
  {Omura}}]{Zhang99:whistlersBS}
\bibinfo{author}{{Zhang}, Y.}, \bibinfo{author}{{Matsumoto}, H.},
  \bibinfo{author}{{Kojima}, H.}, \bibinfo{author}{{Omura}, Y.},
  \bibinfo{year}{1999}.
\newblock \bibinfo{title}{{Extremely intense whistler mode waves near the bow
  shock: Geotail observations}}.
\newblock \bibinfo{journal}{\jgr} \bibinfo{volume}{104},
  \bibinfo{pages}{449--462}.
\newblock \DOIprefix\doi{10.1029/1998JA900049}.
\bibitem[{{Zheng} et~al.(2014){Zheng}, {Chan}, {Albert}, {Elkington}, {Koller},
  {Horne}, {Glauert} and {Meredith}}]{Zheng14:stochastic}
\bibinfo{author}{{Zheng}, L.}, \bibinfo{author}{{Chan}, A.A.},
  \bibinfo{author}{{Albert}, J.M.}, \bibinfo{author}{{Elkington}, S.R.},
  \bibinfo{author}{{Koller}, J.}, \bibinfo{author}{{Horne}, R.B.},
  \bibinfo{author}{{Glauert}, S.A.}, \bibinfo{author}{{Meredith}, N.P.},
  \bibinfo{year}{2014}.
\newblock \bibinfo{title}{{Three-dimensional stochastic modeling of radiation
  belts in adiabatic invariant coordinates}}.
\newblock \bibinfo{journal}{Journal of Geophysical Research (Space Physics)}
  \bibinfo{volume}{119}, \bibinfo{pages}{7615--7635}.
\newblock \DOIprefix\doi{10.1002/2014JA020127}.
\bibitem[{{Zheng} et~al.(2019){Zheng}, {Chen} and {Zhu}}]{Zheng19:emic}
\bibinfo{author}{{Zheng}, L.}, \bibinfo{author}{{Chen}, L.},
  \bibinfo{author}{{Zhu}, H.}, \bibinfo{year}{2019}.
\newblock \bibinfo{title}{{Modeling Energetic Electron Nonlinear Wave-Particle
  Interactions With Electromagnetic Ion Cyclotron Waves}}.
\newblock \bibinfo{journal}{Journal of Geophysical Research (Space Physics)}
  \bibinfo{volume}{124}, \bibinfo{pages}{3436--3453}.
\newblock \DOIprefix\doi{10.1029/2018JA026156}.
\bibitem[{{Zimbardo} and {Perri}(2013)}]{Zimbardo&Perri13}
\bibinfo{author}{{Zimbardo}, G.}, \bibinfo{author}{{Perri}, S.},
  \bibinfo{year}{2013}.
\newblock \bibinfo{title}{{From L{\'e}vy Walks to Superdiffusive Shock
  Acceleration}}.
\newblock \bibinfo{journal}{\apj} \bibinfo{volume}{778}, \bibinfo{pages}{35}.
\newblock \DOIprefix\doi{10.1088/0004-637X/778/1/35}.
\bibitem[{{Zimbardo} and {Perri}(2020)}]{Zimbardo&Perri20}
\bibinfo{author}{{Zimbardo}, G.}, \bibinfo{author}{{Perri}, S.},
  \bibinfo{year}{2020}.
\newblock \bibinfo{title}{{Non-Markovian Pitch-angle Scattering as the Origin
  of Particle Superdiffusion Parallel to the Magnetic Field}}.
\newblock \bibinfo{journal}{\apj} \bibinfo{volume}{903}, \bibinfo{pages}{105}.
\newblock \DOIprefix\doi{10.3847/1538-4357/abb951}.
\bibitem[{{Zimbardo} et~al.(2017){Zimbardo}, {Perri}, {Effenberger} and
  {Fichtner}}]{Zimbardo17}
\bibinfo{author}{{Zimbardo}, G.}, \bibinfo{author}{{Perri}, S.},
  \bibinfo{author}{{Effenberger}, F.}, \bibinfo{author}{{Fichtner}, H.},
  \bibinfo{year}{2017}.
\newblock \bibinfo{title}{{Fractional Parker equation for the transport of
  cosmic rays: steady-state solutions}}.
\newblock \bibinfo{journal}{\aap} \bibinfo{volume}{607}, \bibinfo{pages}{A7}.
\newblock \DOIprefix\doi{10.1051/0004-6361/201731179}.

\end{thebibliography}

\end{document}